\documentclass[lettersize,journal]{IEEEtran}

\usepackage{amsmath,amsfonts}
\usepackage{algorithmic}
\usepackage{algorithm}
\usepackage{array}
\usepackage{adjustbox} 
\usepackage{graphicx} 
\usepackage{subcaption} 
\usepackage{booktabs}
\usepackage[table,xcdraw]{xcolor}
\usepackage{multirow}
\usepackage{amssymb} 
\usepackage{bm} 
\usepackage{enumitem} 
\usepackage{makecell}
\usepackage{hyperref}
\usepackage{soul} 
\usepackage{stfloats}
\usepackage{url}
\usepackage{verbatim}
\usepackage{cite}
\usepackage{array} 
\usepackage{textcomp}
\usepackage{xcolor} 
\usepackage[table,xcdraw]{xcolor}
\usepackage[normalem]{ulem}
\useunder{\uline}{\ul}{}
\usepackage{etoolbox}

\begin{document}

\title{RLHGNN: Reinforcement Learning-driven Heterogeneous Graph Neural Network for Next Activity Prediction in Business Processes}

\author{Jiaxing Wang, Yifeng Yu, Jiahan Song, Bin Cao, Jing Fan and Ji Zhang 
\thanks{*Jing Fan and Ji Zhang are corresponding authors.}
\thanks{Jiaxing Wang, Yifeng Yu, Jiahan Song, Bin Cao, and Jing Fan are with the College of Computer Science and Technology, Zhejiang University of Technology, 310023, Hangzhou, China, and also with Zhejiang Key Laboratory of Visual Information Intelligent Processing, 310023, Hangzhou, China (e-mail: wjx@zjut.edu.cn, yuyifeng@zjut.edu.cn, songjasonjh@gmail.com, fanjing@zjut.edu.cn, bincao@zjut.edu.cn)}
\thanks{Ji Zhang is with the University of Southern Queensland, Toowoomba, QLD 4350, Australia (e-mail: ji.zhang@unisq.edu.au)}
}



\maketitle
\begin{abstract}
Next activity prediction represents a fundamental challenge for optimizing business processes in service-oriented architectures such as microservices environments, distributed enterprise systems, and cloud-native platforms, which enables proactive resource allocation and dynamic service composition. Despite the prevalence of sequence-based methods, these approaches fail to capture non-sequential relationships that arise from parallel executions and conditional dependencies. Even though graph-based approaches address structural preservation, they suffer from homogeneous representations and static structures that apply uniform modeling strategies regardless of individual process complexity characteristics. To address these limitations, we introduce RLHGNN, a novel framework that transforms event logs into heterogeneous process graphs with three distinct edge types grounded in established process mining theory. Our approach creates four flexible graph structures by selectively combining these edges to accommodate different process complexities, and employs reinforcement learning formulated as a Markov Decision Process to automatically determine the optimal graph structure for each specific process instance. RLHGNN then applies heterogeneous graph convolution with relation-specific aggregation strategies to effectively predict the next activity. This adaptive methodology enables precise modeling of both sequential and non-sequential relationships in service interactions. Comprehensive evaluation on six real-world datasets demonstrates that RLHGNN consistently outperforms state-of-the-art approaches. Furthermore, it maintains an inference latency of approximately 1 ms per prediction, representing a highly practical solution suitable for real-time business process monitoring applications. The source code is available at \url{https://github.com/Joker3993/RLHGNN}.
\end{abstract}

\begin{IEEEkeywords}
Next Activity Prediction, Process Mining, Business Process Management, Service-Oriented Computing.

\end{IEEEkeywords}

\section{Introduction}

Service-oriented architectures have fundamentally transformed modern business process implementation, which enables distributed services to coordinate through well-defined interfaces for delivering substantial business value~\cite{liu2023discovering,beheshti2023processgpt}. Within this paradigm, accurately predicting the next activity in running process instances has become essential for enabling proactive resource allocation before service bottlenecks emerge~\cite{marques2024proactive}, facilitating dynamic service composition by anticipating required service types, and supporting real-time decision making within service orchestration engines~\cite{polato2018time}. For instance, in e-commerce platforms, predicting the next activity in order fulfillment processes allows systems to pre-allocate inventory and shipping resources before demand spikes occur. Similarly, in healthcare service workflows, anticipating subsequent diagnostic activities enables hospitals to optimize equipment scheduling and reduce patient wait times. The ability to forecast subsequent service invocations directly impacts system performance, resource utilization efficiency, and overall service quality in contemporary microservices environments.

The fundamental challenge in next activity prediction lies in the inherent diversity of service execution patterns that demand adaptive modeling approaches capable of handling varying complexity levels. Consider a standard e-commerce order fulfillment process where routine purchases generate straightforward sequential service invocations from payment validation through inventory verification to shipping coordination, which follows predictable linear patterns characteristic of well-orchestrated service chains. In contrast, complex B2B procurement scenarios that involve custom specifications produce intricate service interaction patterns with multiple negotiation rounds, document revision services, parallel approval workflows, and conditional branching based on business rules. This diversity reveals a significant limitation in current prediction methodologies - they lack adaptive mechanisms that can dynamically adjust modeling complexity to match individual service orchestration characteristics while maintaining computational efficiency and prediction accuracy.


Traditional approaches to next activity prediction, including rule-based techniques~\cite{sola2021rule}, probabilistic methods~\cite{bohmer2018probability,lakshmanan2015markov}, and machine learning classifiers~\cite{appice2019leveraging,tama2019empirical}, have proven inadequate for handling the complexity and scale of modern service-oriented environments. These methods rely on restrictive assumptions about process structure and require manual feature engineering, which limits their adaptability to diverse business scenarios. As business processes have evolved to incorporate more sophisticated service interactions and real-time requirements, the need for more powerful modeling approaches has become apparent. Current deep learning approaches to next activity prediction follow two primary paradigms, each still presenting distinct limitations when addressing complexity diversity in service orchestrations. Sequence-based methods, which encompass Recurrent Neural Networks (RNNs)\cite{tax2017predictive,camargo2019learning}, convolutional architectures\cite{sun2024next}, transformer-based models~\cite{bukhsh2021processtransformer,jalayer2022ham,nguyen2024switch,zare2025innovative}, and specialized enhancement techniques~\cite{chen2022multi,donadello2023knowledge,theis2022improving,razo2023adjacency}, model process executions as temporally ordered event sequences. Even though these approaches demonstrate superior performance in capturing temporal dependencies and complex non-linear relationships, they fundamentally transform intricate service interactions into sequential representations, potentially discarding important structural relationships that extend beyond temporal ordering, including parallel service executions, and bidirectional service influences characteristic of sophisticated microservices architectures.

To overcome the sequential modeling limitations of traditional approaches, graph-based methods address structural preservation limitations by maintaining service interaction information through network representations~\cite{venugopal2021comparison,rama2023embedding,chiorrini2021exploiting,chiorrini2023multi}. Recent advances have demonstrated effectiveness in capturing complex service dependencies through various graph construction strategies. Despite these improvements, these approaches face a fundamental limitation - they employ uniform graph structures applied identically across all process instances, regardless of individual complexity characteristics. This static modeling strategy operates under the assumption that all service orchestrations benefit from identical representational complexity, which leads to suboptimal outcomes where simple graph structures may inadequately represent complex service dependencies while sophisticated structures introduce unnecessary computational overhead for straightforward service chains.

Generally speaking, there are three major limitations in the existing methods that directly impact service orchestration effectiveness. First, existing methods apply uniform modeling strategies regardless of process complexity diversity, resulting in over-engineered solutions for simple workflows that waste computational resources and under-specified representations for complex orchestrations that miss important structural relationships. Second, current approaches lack automatic adaptation mechanisms to determine appropriate structural representations for different service interaction patterns, forcing practitioners to manually configure graph structures without systematic guidance for optimization. Third, static selection strategies ignore the varying computational requirements of different service compositions in dynamic environments where service discovery and composition decisions must be made within milliseconds to meet real-time performance constraints.

To address these limitations, 
We present in this paper \textbf{RLHGNN} (\textbf{\underline{R}}einforcement \textbf{\underline{L}}earning-driven \textbf{\underline{H}}eterogeneous \textbf{\underline{G}}raph \textbf{\underline{N}}eural \textbf{\underline{N}}etwork), a novel framework using adaptive service interaction modeling with dynamic complexity adjustment. Our approach operates on the principle that different service orchestration instances require varying levels of modeling sophistication, and that optimal graph structure selection can be learned automatically through reinforcement learning (RL) mechanisms rather than relying on static, predetermined graph structures. 
The framework establishes three semantically distinct edge types grounded in established process mining theory: forward edges capturing sequential service dependencies based on directly-follows relationships~\cite{van2016process}, backward edges representing bidirectional service influences derived from dependency analysis studies~\cite{weijters2006process}, and repetition edges modeling iterative service patterns identified through loop detection and recurrence analysis research~\cite{leemans2013discovering}. Based on these three edge types, we construct four adaptive graph structures: forward-only graphs for sequential processes, forward with backward edges for processes requiring historical context, forward with repetition edges for iterative workflows, and comprehensive graphs incorporating all three edge types for complex execution patterns. An RL agent learns to automatically select the most appropriate graph structure for each process instance, and the selected heterogeneous graph structure is then processed using HeteroGraphConv with GraphSAGE aggregators specifically designed for each relationship type, enabling effective modeling of both sequential and non-sequential dependencies for accurate next activity prediction.

Our main contributions provide significant advances in adaptive service orchestration prediction. 
\begin{itemize}
    \item We establish a comprehensive heterogeneous graph structure framework grounded in process mining theory that systematically distinguishes three semantically distinct edge types corresponding to fundamental service interaction relationships, combined with specialized neural processing mechanisms that effectively integrate temporal modeling strengths with structural representation capabilities for service-oriented environments. 
    \item We design an RL-based adaptive selection mechanism formulated as a Markov Decision Process that automatically determines appropriate graph structures for individual service orchestration instances, enabling modeling complexity to dynamically match service interaction characteristics while meeting real-time performance requirements essential for operational deployment.
    \item We demonstrate through comprehensive experimental evaluation that our approach achieves substantial improvements in prediction accuracy across diverse real-world datasets while providing interpretable insights into service execution patterns through selected structures, with particular effectiveness in complex service orchestrations involving parallel execution patterns.
\end{itemize}

The remainder of this paper is organized as follows. Section II reviews related work. Section III provides preliminary concepts and problem formalization. Section IV presents the RLHGNN framework in detail. Section V describes the experimental setup. Section VI presents the performance analysis and results. Section VII concludes this paper.

\section{Related Work}
Over the past decade, extensive research has been conducted to predict the next activity of running business process cases. Traditional approaches encompassed rule-based techniques\cite{sola2021rule}, and  probabilistic methods\cite{bohmer2018probability,lakshmanan2015markov}. Machine learning approaches subsequently emerged, including integrated prediction frameworks\cite{appice2019leveraging}, predictive clustering methods\cite{pravilovic2014process}, decision tree-based techniques for parallel processes\cite{unuvar2016leveraging}, and comprehensive classifier benchmarks\cite{tama2019empirical}. Additional methods utilized extended Markov models with sequence alignment\cite{le2017hybrid} and Naive Bayes techniques for complex process handling\cite{ferilli2019activity}. While these traditional and machine learning methods provided accurate results for simple patterns, they proved limited when handling complex, lengthy process behaviors and required restrictive prior assumptions that constrained model extensibility.

The exponential growth in process data complexity and the need for real-time prediction capabilities in modern service-oriented architectures have necessitated the adoption of deep learning techniques. These methods demonstrate superior performance in capturing intricate non-linear relationships, handling high-dimensional feature spaces, and adapting to dynamic service environments characteristic of contemporary business process management systems. This review focuses specifically on deep learning-based approaches for next activity prediction, categorizing them into sequence-based and graph-based methods. Table~\ref{tab:related_work_comparison} compares existing deep learning approaches across key dimensions.

\begin{table*}[t]
\centering
\caption{Comparison of Existing Next Activity Prediction Methods 
(Methods marked with $\star$ have publicly available code)}
\label{tab:related_work_comparison}
\resizebox{\textwidth}{!}{
\begin{tabular}{|c|c|c|c|c|c|}
\hline
\textbf{Top Category} & \textbf{Sub Category} & \textbf{Method} & \textbf{Year} & \textbf{Core Model} & \textbf{Selection Strategy} \\
\hline

\multirow{17}{*}{\shortstack[c]{\textbf{Sequence-based}\\\textbf{Approaches}}} 
& \multirow{4}{*}{RNN/LSTM-based} 
    & $\star$ Tax et al.\cite{tax2017predictive} & 2017 & LSTM & Static \\
& & $\star$ Camargo et al.\cite{camargo2019learning} & 2019 & LSTM & Static \\
& & Gunnarsson et al.\cite{gunnarsson2023direct} & 2023 & Data-aware LSTM & Static \\
& & Sun et al.\cite{sun2024next} & 2024 & CNN-BiLSTM+Attention & Static \\
\cline{2-6}

& \multirow{5}{*}{Transformer-based}
    & $\star$ ProcessTransformer\cite{bukhsh2021processtransformer} & 2021 & Transformer & Static \\
& & Ni et al.\cite{ni2023predictive} & 2023 & Hierarchical Transformer & Static \\
& & HAM-Net\cite{jalayer2022ham} & 2022 & Hierarchical Attention+LSTM & Static \\
& & Nguyen et al.\cite{nguyen2024switch} & 2024 & Switch-Transformer & Static \\
& & DAW-Transformer\cite{zare2025innovative} & 2025 & Dynamic Transformer & Dataset-level \\
\cline{2-6}

& \multirow{6}{*}{\shortstack[l]{Multi-view, Semantic\\and Alternative}}
    & $\star$ MiTFM\cite{wang2023mitfm} & 2023 & Multi-view Transformer & Static \\
& & $\star$ MiDA\cite{pasquadibisceglie2022multi} & 2022 & Multi-view Deep Learning & Static \\
& & SNAP\cite{oved2025snap} & 2024 & BERT/DeBERTa/GPT-3 & Static \\
& & Chen et al.\cite{chen2022multi} & 2022 & BERT+Transfer Learning & Static \\
& & $\star$ JARVIS\cite{pasquadibisceglie2024jarvis} & 2024 & ViT+Adversarial Training & Static \\
& & $\star$ Donadello et al.\cite{donadello2023knowledge} & 2023 & Attention+Knowledge Modulation & Static \\
\cline{2-6}

& Neural Network Enhancement
    & Theis et al.\cite{theis2022improving} & 2022 & Reachability Graph+NN Masking & Static \\
& & Razo et al.\cite{razo2023adjacency} & 2023 & Adjacency Matrix+Deep Learning & Static \\
\hline

\multirow{7}{*}{\shortstack[c]{\textbf{Graph-based}\\\textbf{Approaches}}}
& \multirow{4}{*}{Homogeneous}
    & Chiorrini et al.\cite{chiorrini2021exploiting}\cite{chiorrini2023multi} & 2021, 2022 & Multi-perspective GNN/Deep GCN & Static \\
& & Rama-Maneiro et al.\cite{rama2023embedding} & 2023 & GCN+RNN Hybrid & Static \\
& & $\star$ SGAP\cite{deng2024enhancing} & 2024 & Sequential Graph+Attention & Static \\
& & $\star$ HiGPP\cite{wang2025higpp} & 2025 & History-informed GNN & Static \\
\cline{2-6}

& Heterogeneous
    & $\star$ MHG-Predictor\cite{wang2025mhg} & 2025 & GatedGraphConv+HeteroSAGE & Static \\
& & \textbf{RLHGNN (Ours)} & 2025 & \textbf{RL+Heterogeneous GNN} & \textbf{Instance-level} \\
\hline

\end{tabular}
}
\end{table*}

\subsection{Sequence-based Approaches}
Sequence-based approaches treat event logs as sequences of activities to capture temporal dependencies in business process execution, enabling effective next activity prediction based on sequential patterns.
\subsubsection{RNN/LSTM-based Approaches}
Early deep learning approaches leveraged recurrent architectures to capture sequential dependencies in business process execution. Tax et al.\cite{tax2017predictive} pioneered Long Short-Term Memory (LSTM)-based prediction using one-hot encoding, while Camargo et al.\cite{camargo2019learning} enhanced accuracy through pre-trained embeddings for categorical attributes. Recent advances include Gunnarsson et al.\cite{gunnarsson2023direct}'s data-aware LSTM networks and Sun et al.\cite{sun2024next}'s hybrid CNN-BiLSTM architectures with self-attention for capturing both local patterns and long-term dependencies.

\subsubsection{Transformer-based Approaches} Transformer architectures have emerged as the dominant paradigm due to their superior long-range dependency modeling capabilities. ProcessTransformer\cite{bukhsh2021processtransformer} demonstrated self-attention superiority for service interaction patterns, while Ni et al.\cite{ni2023predictive} introduced hierarchical transformers for multi-level process structures. HAM-Net\cite{jalayer2022ham} advances this direction through hierarchical attention operating at both attribute and event levels, enabling automatic importance weighting that addresses traditional RNN limitations. Nguyen et al.\cite{nguyen2024switch} introduced Switch-Transformer methodology adapting mixture-of-experts concepts for computational efficiency, while DAW-Transformer\cite{zare2025innovative} represents the most advanced approach with dynamic attribute-aware mechanisms and dataset-level model selection.

\subsubsection{Multi-view, Semantic and Alternative Approaches}
Recent advances focus on multi-view learning and semantic integration. MiDA\cite{pasquadibisceglie2022multi} and MiTFM\cite{wang2023mitfm} introduced multi-view architectures for simultaneous process attribute prediction, while SNAP\cite{oved2025snap} transforms event logs into semantic stories for language model fine-tuning. Chen et al.\cite{chen2022multi} pioneered BERT application through masked activity modeling and multi-task learning frameworks. JARVIS\cite{pasquadibisceglie2024jarvis} combines Vision Transformers with adversarial training for explainable predictions, while Donadello et al.\cite{donadello2023knowledge} presents knowledge-driven modulation systems that integrate procedural knowledge through fuzzy fitness scores for exceptional process executions.

\subsubsection{Neural Network Enhancement Approaches}
Specialized neural network enhancement techniques have emerged to address specific prediction challenges. Theis et al.\cite{theis2022improving} introduced reachability graph-based masking to ensure process model compliance, while Razo et al.\cite{razo2023adjacency} developed adjacency matrix deep learning frameworks for capturing temporal and structural relationships.

Despite strong performance, sequence-based methods face limitations in service environments due to linear processing constraints that struggle with non-sequential relationships essential for parallel service execution patterns. While recent advances introduce dataset-level adaptation, the field lacks instance-level flexibility needed for dynamic service compositions where process complexity varies significantly across execution instances.

\subsection{Graph-based Approaches}
Graph-based approaches model process executions as network structures, enabling capture of complex service interaction patterns while preserving structural relationships that sequence-based approaches might discard.
\subsubsection{Homogeneous Graph Approaches}
Graph-based approaches model process executions as network structures, enabling capture of complex service interaction patterns. Chiorrini et al.\cite{chiorrini2021exploiting}\cite{chiorrini2023multi} developed multi-perspective instance graphs and BIG-DGCNN methodology for next activity prediction using Graph Neural Networks. Rama-Maneiro et al.~\cite{rama2023embedding} integrated GCN with recurrent models for complementary structural and temporal modeling in predictive monitoring, utilizing process discovery techniques to create place graphs from Petri nets.

Recent homogeneous approaches focus on enhanced structural awareness for next activity prediction. SGAP\cite{deng2024enhancing} employs sequential graphs with trace attention mechanisms that dynamically focus on relevant execution paths. HiGPP\cite{wang2025higpp} incorporates history-informed graph construction that explicitly preserves temporal context through specialized edge types.

\subsubsection{Heterogeneous Graph Approaches}
MHG-Predictor~\cite{wang2025mhg} represents significant advancement in heterogeneous graph methods for next activity prediction through multi-layer heterogeneous graphs decomposing process knowledge into four interconnected layers: global process dynamics, instance-level patterns, activities, and contextual attributes. Using GatedGraphConv and HeteroSAGE, it achieved practical deployment in banking systems with 80\% accuracy.

Current graph methods suffer from static construction strategies applied uniformly across instances, ignoring complexity variations critical for efficient service resource allocation. This fails to adapt to varying computational requirements of different service compositions in microservices architectures where service discovery and composition decisions must be made within milliseconds.

\subsection{Positioning of RLHGNN}

RLHGNN addresses three fundamental limitations in existing approaches through corresponding innovations. (1) Current methods apply uniform modeling strategies regardless of process complexity diversity, resulting in over-engineered solutions for simple workflows that waste computational resources and under-specified representations for complex orchestrations that miss important structural relationships. To address this limitation, we establish a comprehensive heterogeneous graph structure framework with four progressive complexity levels that systematically distinguish three semantically distinct edge types corresponding to fundamental service interaction relationships, which advances beyond existing approaches like MHG-Predictor~\cite{wang2025mhg} by enabling precise matching of modeling sophistication to individual process characteristics. (2) Existing approaches lack automatic adaptation mechanisms to determine appropriate structural representations for different service interaction patterns, forcing practitioners to manually configure graph structures without systematic guidance for optimization. We solve this through a RL-based adaptive selection mechanism formulated as a Markov Decision Process that automatically determines the optimal graph structure for each individual process instance, which unlike existing methods that employ predominantly static selection strategies or at best dataset-level adaptation (Table~\ref{tab:related_work_comparison}), provides instance-level flexibility needed for dynamic process environments. Even though DAW-Transformer~\cite{zare2025innovative} attempts dataset-level adaptation, it still lacks the instance-level adaptability our framework provides. (3) Static selection strategies ignore the varying computational requirements of different service compositions in dynamic environments where service discovery and composition decisions must be made within milliseconds to meet real-time performance constraints. Our framework addresses this through specialized neural processing mechanisms that integrate temporal modeling strengths with structural representation capabilities, which enables the RL agent to optimize structure selection based on prediction performance while balancing modeling complexity with computational efficiency, particularly valuable for service-oriented architectures where different service compositions exhibit varying complexities.

\section{Preliminaries and Problem Formulation}
This section introduces the fundamental concepts used throughout the paper and the problem formulation, establishing the theoretical foundation for our RLHGNN framework.

\textbf{Definition 1 (Event).} An event $e = (c, a, t, attr_1, \ldots, attr_m)$ represents an atomic execution of a business activity. Here, $c \in \mathcal{C}$ is the case identifier linking the event to a specific process instance, $a \in \mathcal{A}$ is the activity name from a finite activity set, $t \in \mathbb{T}$ is the timestamp recording when the activity occurred, and $attr_1, \ldots, attr_m$ are additional attributes such as the executing resource, associated cost, or data values. These attributes capture the context essential for understanding process behavior.

\textbf{Definition 2 (Trace).} A trace $\sigma = \langle e_1, e_2, \ldots, e_n \rangle$ represents the complete execution history of a single process instance. Formally, a trace satisfies two properties: (i) case consistency: $\forall e_i, e_j \in \sigma: e_i.c = e_j.c$, ensuring all events belong to the same case; and (ii) temporal ordering: $\forall i < j: e_i.t \leq e_j.t$, preserving the chronological sequence of activities. This temporal ordering is crucial as it captures the control flow logic of the business process.

\textbf{Definition 3 (Prefix Trace).} For a trace $\sigma = \langle e_1, \ldots, e_n \rangle$, the $k$-prefix is defined as $\sigma^{(k)} = \langle e_1, \ldots, e_k \rangle$ where $1 \leq k \leq n$. A prefix trace represents partial execution history and serves as the basis for predicting future activities. The prediction challenge increases as $k$ decreases, since less historical information is available.

\textbf{Definition 4 (Event Log).} An event log $L = \{\sigma_1, \sigma_2, \ldots, \sigma_N\}$ is a multiset of traces, where each trace corresponds to a completed process instance. Event logs capture the actual execution behavior of business processes and serve as the primary data source for process mining and prediction tasks. The multiset notation accounts for the possibility of identical execution sequences occurring multiple times.

\begin{table*}[ht]
\centering
\caption{Example Event Log}
\label{tab:eventlog}
\begin{tabular}{|c|c|c|c|c|}
\hline
\textbf{Event} & \textbf{Case ID} & \textbf{Activity} & \textbf{Timestamp} & \textbf{Resource} \\
\hline
$e_1$ & 1 & SUBMITTED-COMPLETE & 2011/12/25 0:18:09 & 1 \\
$e_2$ & 1 & PARTLYSUBMITTED-COMPLETE & 2011/12/25 0:38:44 & 1 \\
$e_3$ & 1 & PREACCEPTED-COMPLETE & 2011/12/25 0:39:37 & 1 \\
$e_4$ & 1 & Completeren aanvraag-COMPLETE & 2011/12/25 0:39:45 & 2 \\
\hline
\end{tabular}
\end{table*}

Table~\ref{tab:eventlog} exemplifies these concepts using a loan application process. Each row represents an event, with the ``Event'' column showing the event identifier for clarity. Events $e_1$ through $e_4$ form a trace for case 1, capturing the loan application workflow from submission to completion.

\textbf{Definition 5 (Heterogeneous Process Graph).} A heterogeneous process graph $G = (V, E, \mathcal{T}_V, \mathcal{T}_E, \phi, \psi)$ provides a structured representation of process execution patterns. The components are:
\begin{itemize}
    \item $V$: vertex set representing activities in the process
    \item $E \subseteq V \times V$: edge set capturing relationships between activities  
    \item $\mathcal{T}_V = \{\text{activity}\}$: vertex type set (single type in our formulation)
    \item $\mathcal{T}_E = \{\text{forward}, \text{backward}, \text{repeat}\}$: edge type set distinguishing three semantic relationships
    \item $\phi: V \rightarrow \mathcal{T}_V$ and $\psi: E \rightarrow \mathcal{T}_E$: type assignment functions
\end{itemize}

The heterogeneous nature allows differentiated processing of relationships based on their semantic roles in process execution.

\textbf{Definition 6 (Reinforcement Learning Framework).} We formulate graph structure selection as a Markov Decision Process with the following components:
\begin{itemize}
    \item \textbf{State space} $S$: Feature representations of trace prefixes encoding structural complexity
    \item \textbf{Action space} $A = \{0, 1, 2, 3\}$: Four graph structures with increasing complexity
    \item \textbf{Reward function} $R: S \times A \rightarrow \mathbb{R}$: Prediction accuracy-based feedback signal
    \item \textbf{Policy} $\pi: S \rightarrow A$: Learned mapping from trace features to optimal structure
\end{itemize}
This definition enables the agent to learn which graph complexity level best suits each process instance, balancing representational power with computational efficiency.

\textbf{Problem Formulation (Next Activity Prediction).} The next activity prediction problem seeks to estimate the probability distribution over possible subsequent activities given partial execution history. Formally, we define the prediction function as:
\begin{equation}
\Omega_A: G \times \sigma^{(k)} \rightarrow \mathbb{P}(\mathcal{A})
\end{equation}
where $G$ is the selected heterogeneous graph structure for prefix $\sigma^{(k)}$, and $\mathbb{P}(\mathcal{A})$ denotes the probability simplex over the activity set $\mathcal{A}$. The prediction $\hat{a}_{k+1} = \arg\max_{a \in \mathcal{A}} P(a | \sigma^{(k)}, G)$ selects the most probable next activity.

\section{Proposed Approach}
This section presents RLHGNN, an RL-driven heterogeneous graph neural network (HGNN) framework for next activity prediction in business processes. The approach addresses fundamental limitations of existing methods by combining heterogeneous graph construction with adaptive structure selection to capture the diverse complexity patterns inherent in real-world business processes.

\subsection{Framework Overview}

\begin{figure*}
    \centering
    \includegraphics[width=1\textwidth]{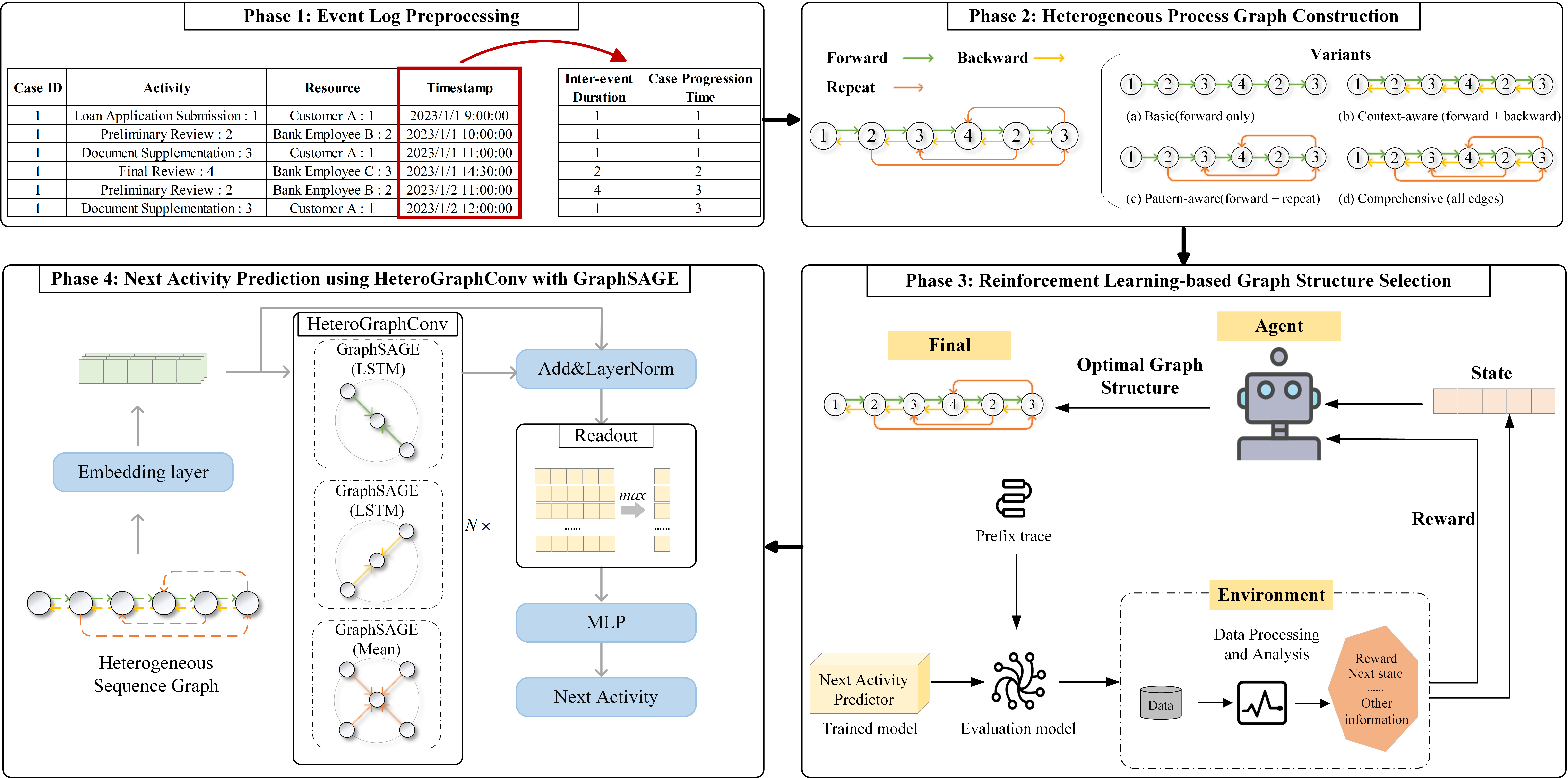}
    \caption{The overall framework of RLHGNN}
    \label{fig:Overall framework}
\end{figure*}

RLHGNN consists of four integrated phases that transform event logs into prediction models, as shown in Fig.~\ref{fig:Overall framework}. The preprocessing phase standardizes the event log and extracts temporal features for each process instance. The heterogeneous process graph construction phase prepares four progressive graph structures with increasing structural complexity as candidate representations for each instance. The RL selection phase dynamically chooses the optimal graph structure for each individual process instance based on learned complexity indicators. Finally, the HGNN processes the selected graph structure to predict the next activity using relation-specific aggregation strategies.

The key idea of our approach is that different process instances require different levels of modeling complexity. Simple sequential processes benefit from basic forward-edge representations, while complex processes with multi-step dependencies and iterative patterns require sophisticated graph structures that capture backward influences and repetition relationships. Rather than applying uniform graph structures across all instances, RLHGNN learns to match structural complexity to process characteristics through RL, automatically determining when historical context access and iterative pattern modeling become critical for prediction accuracy.

\begin{figure}
    \centering
    \includegraphics[width=1\linewidth]{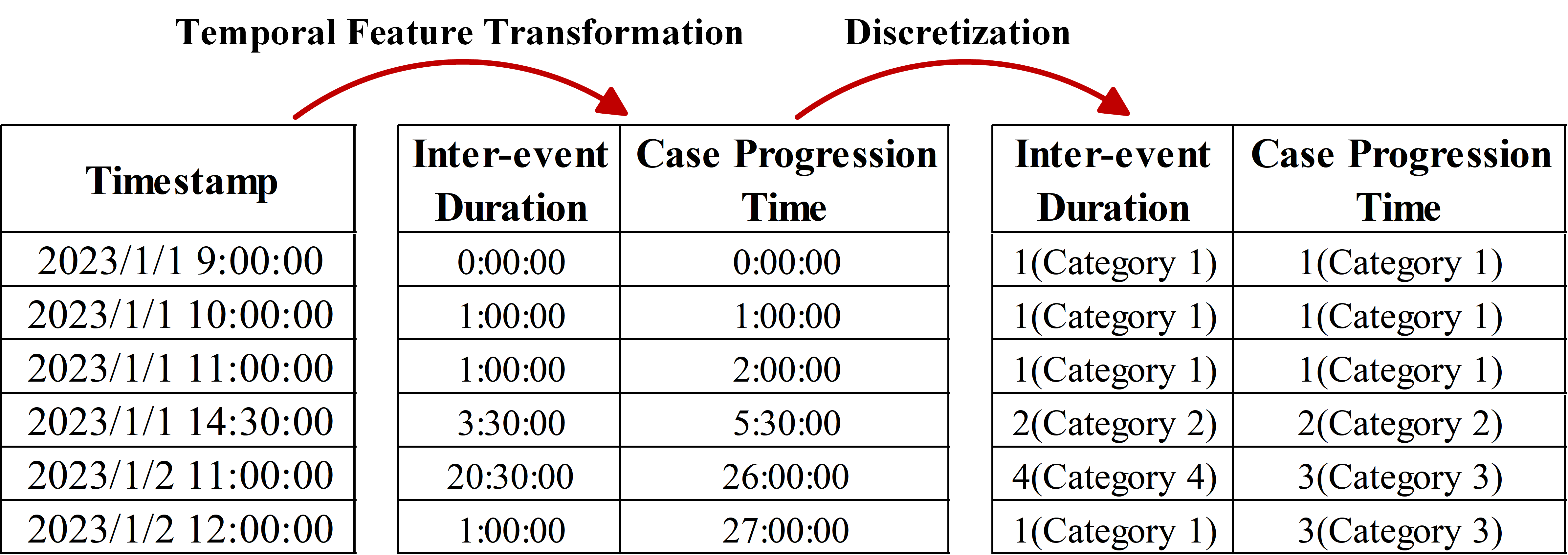}
    \caption{Timestamp processing}
    \label{fig:time_process}
\end{figure}

\subsection{Phase 1: Event Log Preprocessing}


Business process event logs exhibit inherent heterogeneity in attribute types, ranging from categorical activity identifiers and resource assignments to numerical performance metrics and temporal execution timestamps. This diversity creates fundamental incompatibilities for neural network architectures, which require homogeneous input representations with consistent dimensionality and numerical encoding schemes. 

Our framework establishes a unified preprocessing pipeline that systematically transforms heterogeneous attribute types into compatible numerical representations through standardized encoding protocols that preserves  semantic relationships between process elements throughout the transformation process, which proves essential for maintaining prediction accuracy in downstream neural processing stages.

Categorical attributes including activities and resources undergo vocabulary-based encoding constructed from training data. Unknown values in validation and test sets map to reserved tokens, ensuring robustness to novel attributes encountered during deployment. Numerical attributes are discretized based on quantiles, divided into corresponding equal-frequency bins.Taking quartiles as an example, the data is divided into four equal-frequency bins based on quartiles:
\begin{equation}
D(x) = i, \text{ where } q_{i-1} < x \leq q_i \text{ and } i \in \{1,2,3,4\}
\end{equation}
where $q_0, q_1, q_2, q_3, q_4$ represent the 0th, 25th, 50th, 75th, and 100th percentiles, respectively.
This transformation normalizes distributions while preserving relative ordering, crucial for handling diverse numerical scales across attributes.

Temporal features require specialized processing to extract predictive signals. From raw timestamps, we compute inter-event duration $\Delta t_{i,i-1} = t_i - t_{i-1}$ capturing local processing speed, and case progression time $\Delta t_{i,\text{start}} = t_i - t_{\text{start}}$ indicating global lifecycle position. Both undergo quantile discretization, enabling the model to learn temporal patterns at multiple granularities. As illustrated in Fig.~\ref{fig:time_process}, timestamps are transformed into inter-event durations and case progression times, then discretized into categorical values, enabling uniform treatment across all attribute types.

\subsection{Phase 2: Heterogeneous Process Graph Construction}

Traditional homogeneous graph representations treat all process relationships uniformly, failing to capture the semantic distinctions critical for accurate prediction. Building upon Definition 5, we construct heterogeneous process graphs based on the process instances that systematically capture three types of relationships fundamental to business process execution. The construction process creates four graph structures with progressive complexity to enable adaptive modeling based on process characteristics.

\subsubsection{Edge Type Semantics and Construction}

\textbf{Forward edges} capture sequential dependencies by connecting consecutive activities in temporal order. For each process trace $\sigma = \langle e_1, e_2, \ldots, e_n \rangle$, we create directed edges $(v_i, v_{i+1})$ where $v_i$ represents the activity of event $e_i$. These edges form the execution backbone essential for modeling sequential dependencies in business processes, consistent with directly-follows relationships from process discovery literature~\cite{van2016process}.
\textbf{Backward edges} enable access to historical context by connecting each activity to its immediate predecessor~\cite{weijters2006process}. These bidirectional connections facilitate information flow in both temporal directions, crucial for processes where current decisions depend on prior states. The backward edge construction adds edges $(v_{i+1}, v_i)$ for each consecutive activity pair, enabling the model to access historical information during prediction. \textbf{Repeat edges} model iterative patterns by creating bidirectional connections between recurring activity instances~\cite{leemans2013discovering}. The repeat edge construction function \textit{ConnectRepeatedActivities} implements as follows: For each activity $a$ occurring at positions $\{p_1, p_2, \ldots, p_m\}$, we create bidirectional connections by adding both forward edges $(v_{p_i}, v_{p_j+1})$ from earlier occurrences to successors of later occurrences, and backward edges $(v_{p_j}, v_{p_i+1})$ from later occurrences to successors of earlier occurrences, for each pair $(p_i, p_j)$ where $i < j$, provided that the successor positions exist ($p_i + 1 \leq n$ and $p_j + 1 \leq n$). 
This bidirectional design captures the essential characteristic that business process iterations exhibit complex temporal dependencies extending beyond simple forward flow, where decision makers often possess knowledge from the entire process trajectory that influences subsequent choices. 

\begin{figure}
    \centering
    \includegraphics[width=1\linewidth]{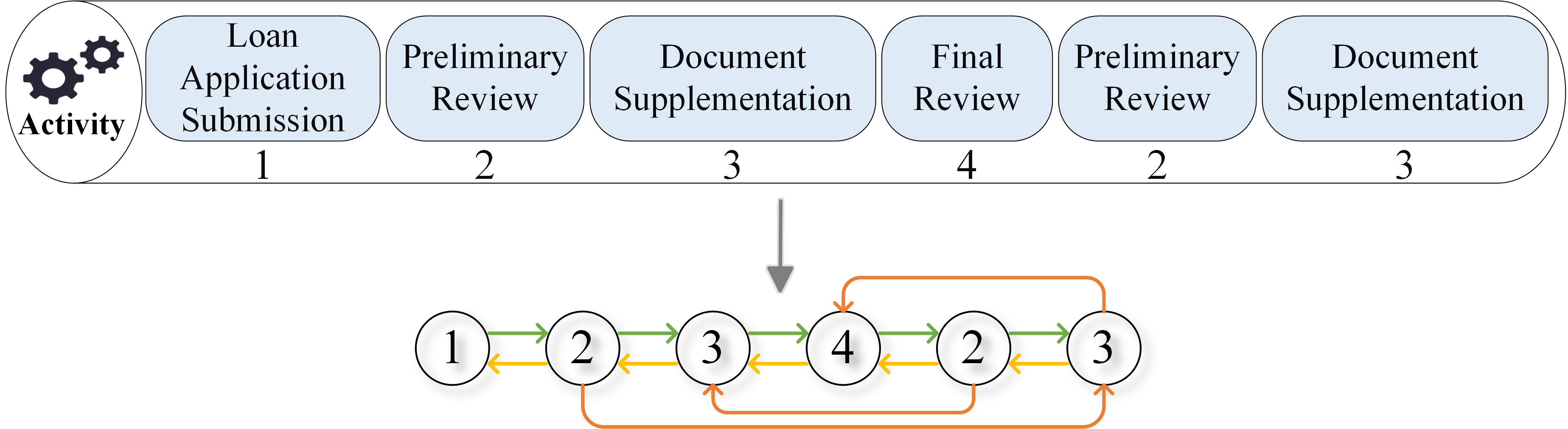}
    \caption{Heterogeneous process graph construction example}
    \label{fig:Hetero_seq_graph}
\end{figure}

In the loan application example shown in Fig.~\ref{fig:Hetero_seq_graph}, activity 2 (Preliminary Review) appears at positions 2 and 5, creating repeat edges from position 2 to position 6 and from position 5 to position 3, enabling the model to learn context-dependent transitions based on iterative review patterns.

\subsubsection{Progressive Graph Structures}

We design four progressive graph structures with increasing representational complexity to enable adaptive modeling:
\begin{itemize}
    \item \textbf{Basic structure $G_1$}: contains only forward edges, suitable for purely sequential processes where activities follow strict temporal order without conditional dependencies or iterative patterns.
    \item \textbf{Context-aware structure $G_2$}: combines forward and backward edges to capture historical dependencies in processes where current activities reference prior states or decisions.
    \item \textbf{Pattern-aware structure $G_3$}: combines forward and repeat edges to model recurring behaviors prevalent in business processes such as document revisions, approval cycles, and iterative development workflows. 
    \item \textbf{Comprehensive structure $G_4$}: combines all three edge types for maximum expressiveness, capturing sequential flow, historical context, and iterative patterns.
\end{itemize} 

\begin{figure}
    \centering
    \includegraphics[width=1\linewidth]{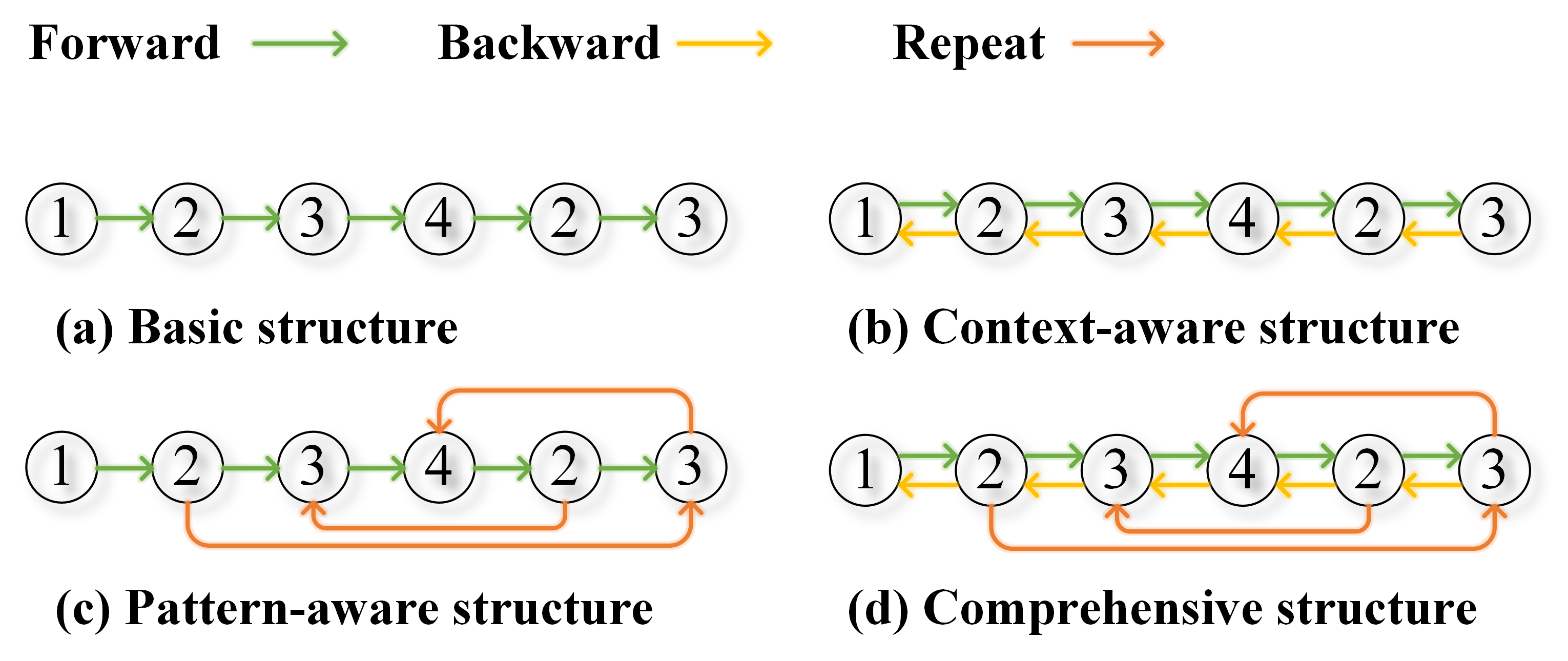}
    \caption{Four progressive graph structures}
    \label{fig:four_graph}
\end{figure}

Fig.~\ref{fig:four_graph} illustrates the four progressive graph structures constructed from the loan application example, and Algorithm~\ref{alg:graph_construction} summarizes the construction process. This progressive design follows the principle of structural parsimony, enabling the RL agent to select minimal complexity sufficient for each process instance while ensuring adequate representational power for complex execution patterns.

\begin{algorithm}[tb]
\caption{Heterogeneous Process Graph Construction}
\label{alg:graph_construction}
\begin{algorithmic}[1]
\STATE \textbf{Input:}  Prefix trace $\sigma^{(k)} = \langle e_1, \ldots, e_k \rangle$
\STATE \textbf{Output:} Four graph structures $\mathcal{G} = \{G_1, G_2, G_3, G_4\}$
\STATE Initialize node set $V \leftarrow \{v_i | v_i \text{ represents activity } e_i.a\}$
\STATE \textbf{Forward edges:} $E_f \leftarrow \{(v_i, v_{i+1}) | i \in [1, k-1]\}$
\STATE \textbf{Backward edges:} $E_b \leftarrow \{(v_{i+1}, v_i) | i \in [1, k-1]\}$
\STATE \textbf{Repeat edges:} $E_r \leftarrow \text{ConnectRepeatedActivities}(\sigma^{(k)})$
\STATE $G_1 \leftarrow (V, E_f)$
\STATE $G_2 \leftarrow (V, E_f \cup E_b)$
\STATE $G_3 \leftarrow (V, E_f \cup E_r)$
\STATE $G_4 \leftarrow (V, E_f \cup E_b \cup E_r)$
\RETURN $\mathcal{G}$
\end{algorithmic}
\end{algorithm}

\subsection{Phase 3: RL-based Graph Structure Selection}
Optimal graph structure selection depends on complex interactions between sequential dependencies, historical context requirements, and iterative patterns that cannot be effectively captured through predetermined heuristic rules. The fundamental challenge lies in determining when a process instance requires backward edges for historical context access, when repeat edges are necessary for iterative pattern modeling, and how to balance representational complexity with computational efficiency. We formulate this selection problem as a Markov Decision Process and employ Deep Q-Networks \cite{mnih2015human} to learn adaptive policies through task-specific optimization.

Traditional rule-based approaches for graph structure selection require domain experts to manually specify selection criteria for determining optimal structural representations, which cannot generalize across diverse process types and fail to adapt to evolving process characteristics. Clustering methods operate under the assumption that similar process instances should utilize identical structural representations, but this assumption breaks down when instances with similar surface characteristics require different modeling approaches due to subtle complexity differences. For example, two loan application processes may exhibit identical activity sequences and similar execution durations, yet one instance may involve straightforward approval workflows that benefit from basic forward-edge modeling, whereas the other may contain complex exception handling with multiple revision cycles that necessitate comprehensive graph structures incorporating backward dependencies and repetition patterns. Recent advances in RL have shown promising results in sequential decision-making and spatial resource allocation~\cite{shakya2023reinforcement}. RL addresses these limitations by learning selection policies directly from task performance, automatically discovering complex feature interactions that correlate with optimal structure choices while balancing immediate computational costs against long-term prediction accuracy.

We formulate graph structure selection as a Markov Decision Process where the state space captures process instance characteristics through five feature categories: structural features including sequence length and unique activity count, temporal features encompassing inter-event duration statistics and rhythm regularity, pattern features identifying repeated activities and loop structures, sequential features analyzing activity transition patterns, and contextual features examining process lifecycle position and resource utilization. The action space $\mathcal{A} = \{G_1, G_2, G_3, G_4\}$ corresponds to our four graph structures, and the reward function balances prediction accuracy with computational efficiency.

\begin{algorithm}[tb]
\caption{RL-based Graph Structure Selection}
\label{alg:rl_structure_selection}
\begin{algorithmic}[1]
\STATE \textbf{Input:} Process prefix $\sigma^{(k)}$, trained Q-network $Q_\theta$
\STATE \textbf{Output:} Selected graph structure $G_i$
\STATE Extract structural features: $f_{struct} \leftarrow [|\sigma^{(k)}|, |\text{unique}(\sigma^{(k)})|, H(\sigma^{(k)})]$
\STATE Extract temporal features: $f_{temp} \leftarrow [\mu(\Delta T), \sigma^2(\Delta T), \text{rhythm\_score}(\Delta T)]$
\STATE Extract pattern features: $f_{pattern} \leftarrow [\text{repeat\_count}(\sigma^{(k)}), \text{loop\_depth}(\sigma^{(k)})]$
\STATE Combine features: $s \leftarrow \text{concat}[f_{struct}, f_{temp}, f_{pattern}]$
\STATE Select structure: $a^* \leftarrow \arg\max_a Q_\theta(s, a)$
\RETURN Graph structure $G_{a^*}$
\end{algorithmic}
\end{algorithm}

The reward function design addresses the core challenge of balancing prediction accuracy with computational efficiency:
\begin{equation}
\begin{split}
\mathcal{R}(s, a) =\ & \alpha \cdot Accuracy(s, a) 
- \beta \cdot ComputeCost(a) \\
& + \gamma \cdot Efficiency(s, a)
\end{split}
\end{equation}

\noindent where $Accuracy(s, a)$ measures prediction performance using the selected structure, $ComputeCost(a)$ penalizes unnecessarily complex structures, and $Efficiency(s, a)$ rewards achieving high accuracy with simpler structures. The hyperparameters $\alpha$, $\beta$, and $\gamma$ control the relative importance of accuracy, computational cost, and efficiency respectively, allowing adaptation to different deployment scenarios with varying computational constraints and performance requirements.

Our DQN implementation employs a three-layer architecture with feature-specific encoders for different feature categories, followed by a fusion layer and Q-value output. Through training across diverse process types, the agent learns to automatically identify when historical context becomes critical, such as in processes with conditional branching where current decisions depend on earlier choices, iterative workflows where previous iteration outcomes influence subsequent actions, and exception handling scenarios where unusual patterns require reference to earlier process states. The learned policy effectively captures these relationships through feature interactions that would be difficult to specify manually, enabling the agent to determine context requirements based on process complexity indicators rather than predetermined rules. Algorithm 2 summarizes the implementation of Phase 3. 

\subsection{Phase 4: Next Activity Prediction using HeteroGraphConv with GraphSAGE}

The selected heterogeneous graph structure from Phase 3 requires specialized neural processing that respects the semantic differences between edge types established in Phase 2. We employ HeteroGraphConv\cite{hamilton2017inductive} with GraphSAGE aggregators specifically designed for each relationship type, enabling effective modeling of both sequential and non-sequential dependencies in business processes.

We selected GraphSAGE as the aggregation mechanism for our HeteroGraphConv framework after careful consideration of alternative GNN architectures. While graph convolutional networks (GCNs) demonstrate strong performance on static graphs, they require the entire graph structure during training, limiting their applicability to dynamic business processes where new activity patterns may emerge. Graph attention networks (GATs) provide learnable attention weights but introduce significant computational overhead, particularly problematic when processing multiple graph structures in our RL framework.

GraphSAGE offers three critical advantages for our application: (1) inductive learning capability that generalizes to previously unseen process instances containing novel activity combinations, essential for real-world deployment where new process variants continuously emerge; (2) sampling-based aggregation with linear time complexity $O(k)$ with respect to the sampling size, enabling efficient processing of large-scale event logs compared to GAT's quadratic complexity; and (3) flexible aggregation functions (LSTM, mean, pool) that can be tailored to different edge types in our heterogeneous graph structure. These characteristics align perfectly with the requirements of adaptive business process prediction, where computational efficiency must be balanced with representational power across diverse process patterns.

Our model implements differentiated aggregation mechanisms tailored to each edge type present in the selected graph structure. Sequential relationships captured by forward and backward edges undergo LSTM-based aggregation to preserve temporal ordering, while pattern relationships captured by repeat edges utilize mean aggregation to capture common characteristics across recurring activity instances:

\begin{align}
h_v^{fwd} &= \text{LSTM}_{fwd}(\{h_u \mid u \in \mathcal{N}_{fwd}(v)\})\\
h_v^{bwd} &= \text{LSTM}_{bwd}(\{h_u \mid u \in \mathcal{N}_{bwd}(v)\})\\
h_v^{rep} &= \text{MEAN}(\{h_u \mid u \in \mathcal{N}_{rep}(v)\})
\end{align}

\noindent where $h_v^{fwd}$, $h_v^{bwd}$, and $h_v^{rep}$ represent the aggregated embeddings for forward, backward, and repeat edge types respectively, $\mathcal{N}_{type}(v)$ denotes the set of neighboring nodes connected to node $v$ via edges of the specified type, and $h_u$ represents the feature vector of neighboring node $u$.

For each graph structure selected by the RL policy, our HGNN dynamically combines aggregation results using structure-specific weight matrices:

\begin{align}
h_v^{(1)} &= \sigma(W_{fwd} \cdot h_v^{fwd} + W_{self} \cdot h_v) \quad \text{(Basic)} \\
h_v^{(2)} &= \sigma(W_{fwd} \cdot h_v^{fwd} + W_{bwd} \cdot h_v^{bwd} + W_{self} \cdot h_v) \quad  \\
h_v^{(3)} &= \sigma(W_{fwd} \cdot h_v^{fwd} + W_{rep} \cdot h_v^{rep} + W_{self} \cdot h_v) \quad  \\
h_v^{(4)} &= \sigma(W_{fwd} \cdot h_v^{fwd} + W_{bwd} \cdot h_v^{bwd} + W_{rep} \cdot h_v^{rep}\nonumber \\
 &\quad + W_{self} \cdot h_v)
\end{align}

\noindent where $\sigma$ represents the ReLU activation function, $W_{fwd}$, $W_{bwd}$, $W_{rep}$, and $W_{self}$ are learnable weight matrices for forward, backward, repeat, and self-connection transformations respectively, and $h_v$ denotes the initial feature vector of node $v$. This differentiated processing strategy corresponds directly to the four graph structures from Phase 2, enabling specific types of information flow based on edge composition.

Business process events contain diverse attributes beyond activity names that significantly influence prediction accuracy. Our framework systematically integrates temporal attributes, resource information, and case-specific metadata through an initial embedding layer that transforms these heterogeneous features into a unified dimensional space. The embedded representations are then processed by the HeteroGraphConv module with relation-specific GraphSAGE aggregators as defined in Equations (4)-(6). To enhance training stability and gradient flow, we employ residual connections with layer normalization (Add\&LayerNorm) after each graph convolution layer. The architecture uses a readout function to aggregate node representations, focusing on the current activity position within the process instance. These aggregated features are then refined through a two-layer MLP with ReLU activation before the final classification layer produces probability distributions over possible next activities:

\begin{equation}
P(a_{next}|G_c) = \text{softmax}(W_{class} \cdot h_{current} + b_{class})
\end{equation}

\noindent where $h_{current}$ represents the embedding of the current node position, and $W_{class}$ and $b_{class}$ are learnable classification parameters optimized jointly with the graph neural network components.

\subsection{Complexity Analysis}

The computational complexity analysis reveals balanced trade-offs between adaptivity and efficiency. Graph construction exhibits time complexity $O(k + m_{\text{total}}^2)$ where $k$ represents trace length and $m_{\text{total}}$ denotes the total number of repeat edge connections across all recurring activities. Forward and backward edge construction requires $O(k)$ operations each, while repeat edge construction necessitates $O(m_i^2)$ operations for each activity $i$ with $m_i$ repetitions, yielding total repeat complexity $O(\sum_i m_i^2)$. GraphSAGE training requires $O(L \cdot |V| \cdot |E_f \cup E_b \cup E_r| \cdot d \cdot s)$ complexity, where $L$ represents network layers, $|V|$ denotes vertex count, $E_f$, $E_b$, $E_r$ represent forward, backward, and repeat edge sets respectively, $d$ indicates embedding dimensionality, and $s$ represents sampling neighborhood size. RL training introduces overhead $O(T \cdot (|S| \cdot d_{\text{net}} + |A| \cdot \text{batch\_size}))$ where $T$ indicates training episodes, $d_{\text{net}}$ represents network computation cost, and batch processing efficiency affects the actual computational requirements.

Traditional static methods incur uniform complexity $\text{Cost}(G_4)$ for all instances, whereas RLHGNN achieves instance-optimized complexity through adaptive selection, resulting in expected complexity $\mathbb{E}[\text{Cost}(G_i)]$ where $i \in \{1,2,3,4\}$ is selected based on process characteristics. For datasets exhibiting structural diversity, this yields $\mathbb{E}[\text{Cost}(G_i)] < \text{Cost}(G_4)$ due to the probability distribution favoring simpler structures for less complex process instances. This theoretical advantage translates directly to deployment efficiency, where straightforward processes utilize basic structures with $O(|V| + |E_f|)$ complexity, while complex orchestrations receive comprehensive structures with $O(|V| + |E_f| + |E_b| + |E_r|)$ complexity only when the adaptive mechanism determines such sophistication necessary.

Compared to static approaches applying uniform structures, RLHGNN achieves superior computational efficiency through adaptive selection mechanisms that scale performance gains proportionally with dataset structural diversity. The framework remains practically viable for large-scale business process prediction scenarios, as the RL training overhead becomes amortized across numerous inference operations. During deployment, structure selection requires only $O(1)$ forward pass through the trained policy network, making the adaptive approach computationally negligible compared to the graph processing costs.

\section{Experimental Setup}

\subsection{Benchmark Event Log Datasets}

We evaluate RLHGNN using six publicly available event logs from the 4TU Research Data Center\footnote{\url{https://data.4tu.nl}}, carefully selected to represent diverse business domains and process characteristics that test our adaptive framework's capabilities across varying structural complexities. Table~\ref{tab:statistics} summarizes the statistical properties of these datasets, which exhibit substantial variation in scale (1,487 to 13,087 cases), complexity (4 to 29 unique activities), and structural patterns (12.2\% to 95.3\% case repetition ratios). 

\begin{table}[htp]
    \centering
    \caption{Statistical Characteristics of Benchmark Event Log Datasets}
    \label{tab:statistics}
    \scalebox{0.9}{
    \begin{tabular}{@{}ccccccc@{}}
        \toprule
        \textbf{Dataset}  & 
        \textbf{\begin{tabular}[c]{@{}c@{}}Case\\Number\end{tabular}} & 
        \textbf{\begin{tabular}[c]{@{}c@{}}Activity\\Number\end{tabular}} & 
        \textbf{\begin{tabular}[c]{@{}c@{}}Event\\Number\end{tabular}} & 
        \textbf{\begin{tabular}[c]{@{}c@{}}Average\\Case\\Length\end{tabular}} & 
        \textbf{\begin{tabular}[c]{@{}c@{}}Max\\Case\\Length\end{tabular}} & 
        \textbf{\begin{tabular}[c]{@{}c@{}}Case\\Repetition\\Ratio\end{tabular}} \\ \midrule
        BPI12C & 13,087 & 23 & 164,506 & 13 & 96 & 53.6\% \\
        BPI12CW & 9,658 & 6 & 72,413 & 7 & 74 & 72.7\% \\
        BPI13I & 7,554 & 13 & 65,533 & 9 & 123 & 95.3\% \\
        BPI13P & 2,306 & 7 & 9,011 & 4 & 35 & 47.2\% \\
        BPI13CP & 1,487 & 4 & 6,660 & 4 & 35 & 65\% \\
        BPI2020P & 2,099 & 29 & 18,246 & 9 & 21 & 12.2\% \\ \bottomrule
    \end{tabular}
    }
\end{table}

The evaluation datasets span three distinct business domains that represent different operational characteristics and complexity patterns. The financial services domain includes BPI12C and BPI12CW\footnote{\url{https://data.4tu.nl/articles/_/12689204/1}}, which represent loan application processes at different granularities. BPI12C captures the complete loan application workflow including assessment, documentation, and approval procedures (23 activities, 13,087 cases). BPI12CW provides a filtered perspective of the same dataset, concentrating on core processing activities (6 activities, 9,658 cases), where activities with frequency below a specified threshold have been removed to create a streamlined process view. Both datasets exhibit moderate to high repetition patterns (53.6\% and 72.7\% respectively), which reflects standardized financial procedures.

The IT service management domain utilizes three datasets from Volvo IT's incident and problem management system, each representing different aspects of service operations. BPI13I\footnote{\url{https://data.4tu.nl/articles/_/12693914/1}} encompasses comprehensive incident lifecycle management (13 activities, 7,554 cases) with the highest repetition ratio (95.3\%), which indicates highly standardized procedures. BPI13P\footnote{\url{https://data.4tu.nl/datasets/7aafbf5b-97ae-48ba-bd0a-4d973a68cd35/1}} focuses on problem reporting and handling (7 activities, 2,306 cases) with moderate repetition (47.2\%), whereas BPI13CP\footnote{\url{https://data.4tu.nl/datasets/1987a2a6-9f5b-4b14-8d26-ab7056b17929}} covers closed problem resolution workflows (4 activities, 1,487 cases) with 65.0\% repetition. These datasets reflect varying process maturity levels within IT service operations, providing different structural challenges for our adaptive selection mechanism.

The administrative domain features BPI2020P\footnote{\url{https://data.4tu.nl/datasets/fb84cf2d-166f-4de2-87be-62ee317077e5/1}}, which represents administrative processes through university travel reimbursement workflows. This dataset exhibits the highest complexity (29 activities) with the lowest repetition ratio (12.2\%), which reflects the ad-hoc nature of administrative procedures involving multiple approval levels and exception handling. This dataset provides a particularly challenging test case for our adaptive framework due to its high variability and complex approval patterns.

\subsection{Experimental Setup}

We employ three-fold cross-validation following established protocols in next activity prediction literature to ensure direct comparability with existing methods. Each fold uses an 80-20 training-validation split, with training data further divided equally between baseline establishment and RL agent training.

The HGNN implements a two-layer HeteroGraphConv architecture with GraphSAGE aggregators, containing 128 hidden units per layer. Training uses NAdam optimization (learning rate: 0.001), dropout regularization (0.1), and 64-sample mini-batches. The RL component employs a three-layer Deep Q-Network [256, 128, 128] with ReLU activations, Adam optimization (learning rate: 0.0001), 50,000-transition experience replay buffer, epsilon-greedy exploration (1.0 to 0.1 decay), target network updates every 2,000 steps, and discount factor $\gamma = 0.99$.

Event logs undergo preprocessing through Pm4Py with quantile-based discretization for temporal attributes. Activity embeddings match unique activity counts per dataset. Training extends to 100 epochs with early stopping after 10 consecutive epochs without validation accuracy improvement.

\subsection{Evaluation Metrics}
For baseline analysis, we employ accuracy and macro F1-score as primary metrics, consistent with established practice in imbalanced multi-class prediction tasks\cite{pasquadibisceglie2022multi}. The macro F1-score is computed as:
\begin{equation}
\text{F1-score}_{\text{macro}} = \frac{1}{|A|} \sum_{a \in A} \frac{2 \cdot \text{precision}_a \cdot \text{recall}_a}{\text{precision}_a + \text{recall}_a}
\end{equation}
where $A$ represents the activity set. 

In ablation studies, we additionally report the geometric mean (GMean) of sensitivity and specificity\cite{pasquadibisceglie2024jarvis}:
\begin{equation}
\text{GMean} = \sqrt{\text{sensitivity} \times \text{specificity}}
\end{equation}
This metric provides balanced assessment for datasets with severe class imbalance. All implementations use Python 3.9 with PyTorch 2.0.1 on identical hardware configurations.

\section{Performance Analysis}
This section evaluates RLHGNN through comprehensive experiments on the benchmark event logs datasets. We structure our evaluation around four key research questions that examine different aspects of our approach:

\textbf{RQ1}: How does RLHGNN's adaptive graph structure selection compare to state-of-the-art static methods in terms of prediction accuracy and F1-score across diverse business process datasets?

\textbf{RQ2}: What is the individual contribution of each component in RLHGNN, and how does adaptive structure selection compare to using fixed graph structures?

\textbf{RQ3}: How does RLHGNN's computational overhead compare to baseline methods in terms of training efficiency and inference latency?

\textbf{RQ4}: What is the effectiveness of RLHGNN's heterogeneous aggregation approach compared to uniform aggregation strategies?

\subsection{Baseline Analysis} 
To answer Q1, we compare RLHGNN with six state-of-the-art methods selected from Table \ref{tab:related_work_comparison}, including sequence-based approaches (MiDA \cite{pasquadibisceglie2022multi}, MiTFM \cite{wang2023mitfm}), graph-based methods (HiGPP \cite{wang2025higpp}, SGAP \cite{deng2024enhancing}, JARVIS \cite{pasquadibisceglie2024jarvis}, MHG-Predictor \cite{wang2025mhg}). To ensure fair and reproducible comparisons, baseline methods were selected based on three criteria: (1) publication within the past five years, (2) appearance in reputable conferences or journals, and (3) availability of publicly accessible implementations. All methods were evaluated under identical experimental conditions, including consistent data preprocessing pipelines, hardware configurations, and hyperparameter optimization procedures to ensure fair comparison.

Table \ref{tab:baseline} demonstrates RLHGNN's performance advantages across evaluation metrics. RLHGNN achieves the highest average performance in both accuracy (0.782) and F1-score (0.578), with particularly pronounced improvements on complex datasets. The performance patterns reveal the technical advantages of RLHGNN's instance-level adaptive approach. On BPI12CW, RLHGNN outperforms the strongest baseline (HiGPP) by 3.4\% in accuracy and 3.8\% in F1-score through its ability to dynamically match graph structure complexity to individual process instance requirements, selecting appropriate structures based on each instance's sequence length, activity patterns, and temporal dependencies rather than applying uniform configurations.

RLHGNN exhibits limitations on extremely sparse datasets that reveal important insights about adaptive approaches. On BPI2020P, RLHGNN underperforms compared to MHG-Predictor in accuracy (0.873 vs 0.886), highlighting scenarios where adaptive structure selection becomes counterproductive. This dataset's extreme sparsity (29 activities across 2,099 cases with 12.2\% repetition ratio) creates insufficient training signal for effective policy learning, leading to suboptimal structure choices. The results indicate that adaptive approaches require sufficient structural regularity to function effectively, while static methods may provide better generalization in highly sparse domains.

Traditional graph-based methods demonstrate stable but static performance, while sequence-based approaches exhibit inconsistent results across datasets, highlighting the importance of dynamic adaptation to varying process characteristics. The consistent improvements achieved by RLHGNN stem from adaptive structure selection preventing computational waste, relation-specific aggregation optimally utilizing edge semantics, and learned policies capturing feature interactions that resist manual specification.

\begin{table*}[ht]
\centering
\caption{Performance Comparison of RLHGNN Against Baseline Methods}
\label{tab:baseline}
\begin{tabular}{cccccccccc}
\toprule
\textbf{Metrics} & \textbf{Method Type} & \textbf{Method} & \textbf{BPI12C} & \textbf{BPI12CW} & \textbf{BPI13I} & \textbf{BPI13P} & \textbf{BPI13CP} & \textbf{BPI2020P} & \textbf{Average} \\
\midrule

\multirow{8}{*}{Accuracy} 
& \multirow{2}{*}{Sequence-based} & MiDA           & 0.782 & 0.839 & 0.732 & 0.625 & 0.652 & 0.867 & 0.749 \\
&                                  & MiTFM          & 0.800 & 0.842 & 0.743 & 0.636 & 0.641 & 0.884 & 0.758 \\
\cmidrule(l){2-10}
& \multirow{5}{*}{Graph-based}     & HiGPP          & 0.800 & 0.853 & 0.761 & 0.657 & 0.683 & 0.881 & 0.772 \\
&                                  & JARVIS         & 0.781 & 0.823 & 0.664 & 0.488 & 0.524 & 0.831 & 0.685 \\
&                                  & SGAP           & 0.719 & 0.787 & 0.598 & 0.506 & 0.567 & 0.851 & 0.671 \\
&                                  & MHG-Predictor  & 0.799 & 0.874 & 0.721 & 0.649 & 0.674 & \textbf{\underline{0.886}} & 0.767 \\
&                                  & \textbf{RLHGNN (ours)} & \textbf{\underline{0.806}} & \textbf{\underline{0.887}} & \textbf{\underline{0.764}} & \textbf{\underline{0.667}} & \textbf{\underline{0.693}} & 0.873 & \textbf{\underline{0.782}} \\
\midrule

\multirow{8}{*}{F1-score} 
& \multirow{2}{*}{Sequence-based} & MiDA           & 0.642 & 0.715 & 0.437 & 0.435 & 0.456 & 0.564 & 0.542 \\
&                                  & MiTFM          & 0.662 & 0.714 & 0.431 & 0.414 & 0.437 & 0.571 & 0.539 \\
\cmidrule(l){2-10}
& \multirow{5}{*}{Graph-based}     & HiGPP          & 0.664 & 0.755 & 0.423 & 0.435 & 0.468 & 0.562 & 0.551 \\
&                                  & JARVIS         & 0.621 & 0.693 & 0.351 & 0.418 & 0.452 & 0.474 & 0.502 \\
&                                  & SGAP           & 0.518 & 0.635 & 0.216 & 0.201 & 0.302 & 0.468 & 0.390 \\
&                                  & MHG-Predictor  & 0.668 & 0.774 & 0.391 & 0.416 & \textbf{\underline{0.512}} & 0.561 & 0.554 \\
&                                  & \textbf{RLHGNN (ours)} & \textbf{\underline{0.684}} & \textbf{\underline{0.793}} & \textbf{\underline{0.449}} & \textbf{\underline{0.464}} & 0.491 & \textbf{\underline{0.584}} & \textbf{\underline{0.578}} \\
\bottomrule
\end{tabular}
\end{table*}

\begin{table*}[]
    \centering
    \makeatletter
    \patchcmd{\@makecaption}{\scshape}{}{}{}
    \caption{Ablation Study Results of RLHGNN}
    \makeatother

\begin{tabular}{@{}ccccccccc@{}}
\toprule
\textbf{Metrics}                   & \textbf{Graph Structure} & \textbf{BPI12C}      & \textbf{BPI12CW}     & \textbf{BPI13CP}     & \textbf{BPI13I}      & \textbf{BPI13P}      & \textbf{BPI2020P}    & \textbf{Average}        \\ \midrule
\multirow{5}{*}{\textbf{GMean}}    & Basic            & 0.798                & 0.856                & 0.640                & 0.622                & 0.640                & 0.645                & 0.700                \\
                                   & Context-aware    & 0.804                & {\ul \textbf{0.876}} & 0.636                & 0.651                & {\ul \textbf{0.659}} & 0.702                & 0.721                \\
                                   & Pattern-aware    & 0.798                & 0.849                & 0.634                & 0.628                & 0.644                & 0.661                & 0.702                \\
                                   & Comprehensive    & 0.804                & 0.859                & 0.636                & 0.655                & 0.658                & 0.739                & 0.725                \\
                                   & \textbf{RLHGNN}  & {\ul \textbf{0.806}} & 0.872                & {\ul \textbf{0.654}} & {\ul \textbf{0.657}} & 0.656                & {\ul \textbf{0.743}} & {\ul \textbf{0.731}} \\ \midrule
\multirow{5}{*}{\textbf{F1-score}} & Basic            & 0.670                & 0.777                & 0.475                & 0.395                & 0.447                & 0.412                & 0.529                \\
                                   & Context-aware    & 0.678                & {\ul \textbf{0.797}} & 0.475                & 0.433                & {\ul \textbf{0.469}} & 0.503                & 0.559                \\
                                   & Pattern-aware    & 0.671                & 0.756                & 0.479                & 0.407                & 0.451                & 0.435                & 0.533                \\
                                   & Comprehensive    & 0.678                & 0.780                & 0.478                & 0.438                & 0.467                & 0.568                & 0.568                \\
                                   & \textbf{RLHGNN}  & {\ul \textbf{0.684}} & 0.784                & {\ul \textbf{0.491}} & {\ul \textbf{0.449}} & 0.464                & {\ul \textbf{0.584}} & {\ul \textbf{0.576}} \\ \bottomrule
\end{tabular}

\label{tab:ablation}

\end{table*}

\subsection{Ablation Study}

To answer Q2, we conduct an ablation study by removing the RL module and training separate models with each of the four fixed graph structures: Basic (forward edges only), Context-aware (forward and backward edges), Pattern-aware (forward and repeat edges), and Comprehensive (all three edge types). This systematic evaluation reveals how adaptive structure selection contributes to performance. 

Table \ref{tab:ablation} demonstrates that RLHGNN with dynamic structure selection consistently outperforms all static structures in terms of average effectiveness performance on all datasets, achieving an average GMean of 0.731 and F1-score of 0.576 across six datasets. The performance patterns reveal important insights about matching graph structures to process characteristics. The basic graph structures show varying effectiveness across datasets, achieving strong performance on some datasets like BPI12CW (GMean: 0.856) while experiencing significant degradation on others such as BPI13I (GMean: 0.622), confirming that forward edges alone cannot capture non-sequential dependencies. The Context-aware structures demonstrate consistent improvements across most datasets, particularly evident on BPI2020P where GMean increases from 0.645 to 0.702, indicating the value of backward edge modeling for capturing historical dependencies. The Pattern-aware graph structure yields mixed results, suggesting that explicit pattern modeling benefits specific process types rather than providing universal improvements. The Comprehensive graph structure achieves strong performance but does not uniformly outperform simpler structures, validating our core hypothesis that optimal complexity varies by process instance. RLHGNN's adaptive mechanism enables it to select appropriate structures dynamically, achieving consistent gains without manual graph structure.

\begin{figure}
    \centering
    \begin{subfigure}[b]{0.24\textwidth}
        \includegraphics[trim=0 0 20 0, clip,width=\textwidth]{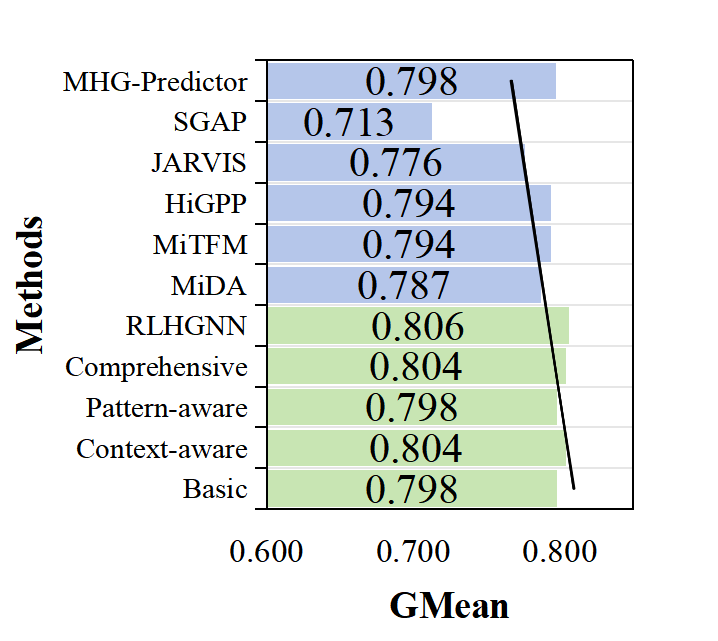}
        \caption{BPI12C}
        \label{BPI12C}
    \end{subfigure}
    \hfill
    \begin{subfigure}[b]{0.24\textwidth}
        \includegraphics[trim=0 0 20 0, clip,width=\textwidth]{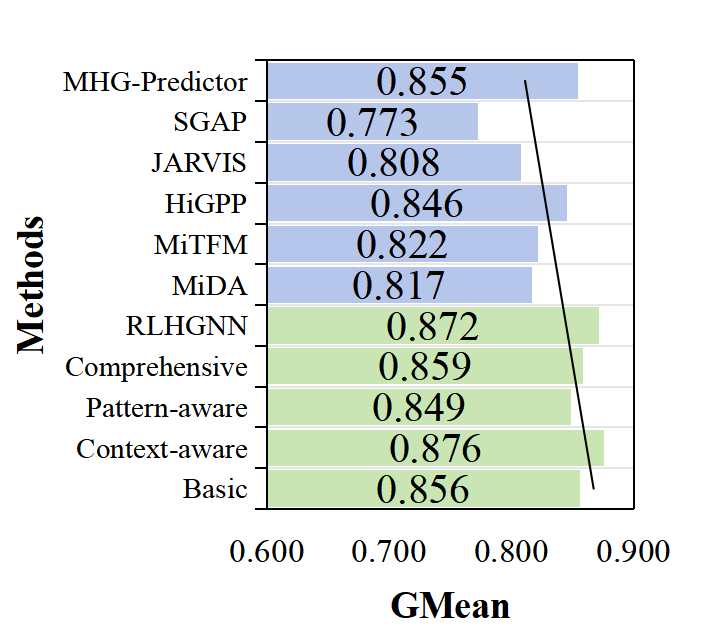}
        \caption{BPI12CW}
        \label{BPI12CW}
    \end{subfigure}
    
    \begin{subfigure}[b]{0.24\textwidth}
        \includegraphics[trim=0 0 20 0, clip,width=\textwidth]{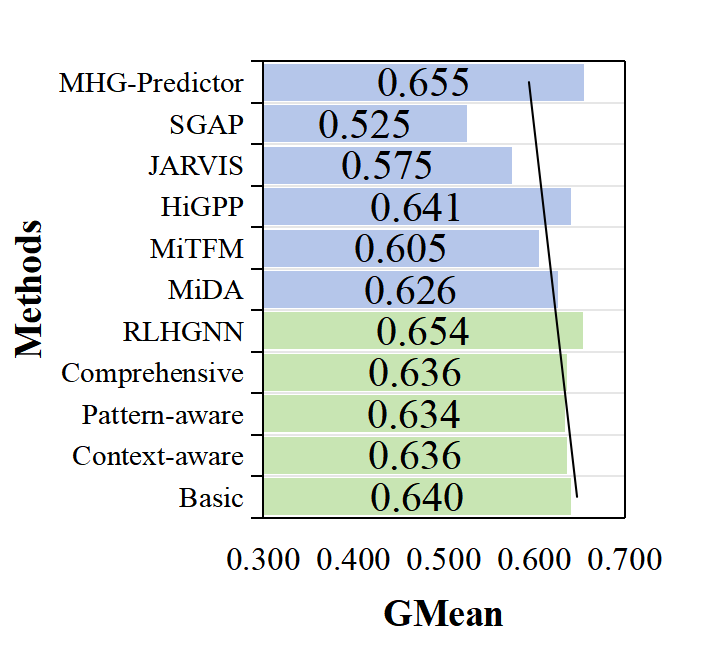}
        \caption{BPI13CP}
        \label{BPI13CP}
    \end{subfigure}
    \hfill
    \begin{subfigure}[b]{0.24\textwidth}
        \includegraphics[trim=0 0 20 0, clip,width=\textwidth]{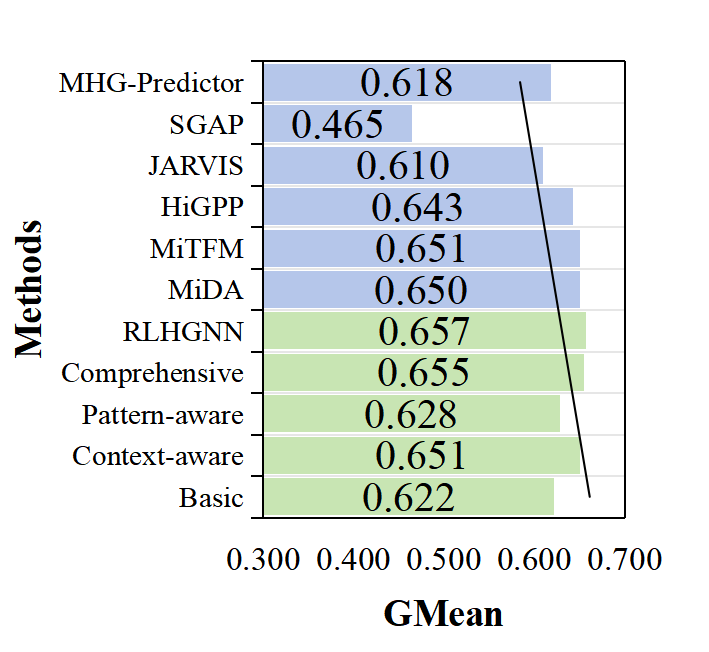}
        \caption{BPI13I}
        \label{BPI13I}
    \end{subfigure}

    \begin{subfigure}[b]{0.24\textwidth}
        \includegraphics[trim=0 0 20 0, clip,width=\textwidth]{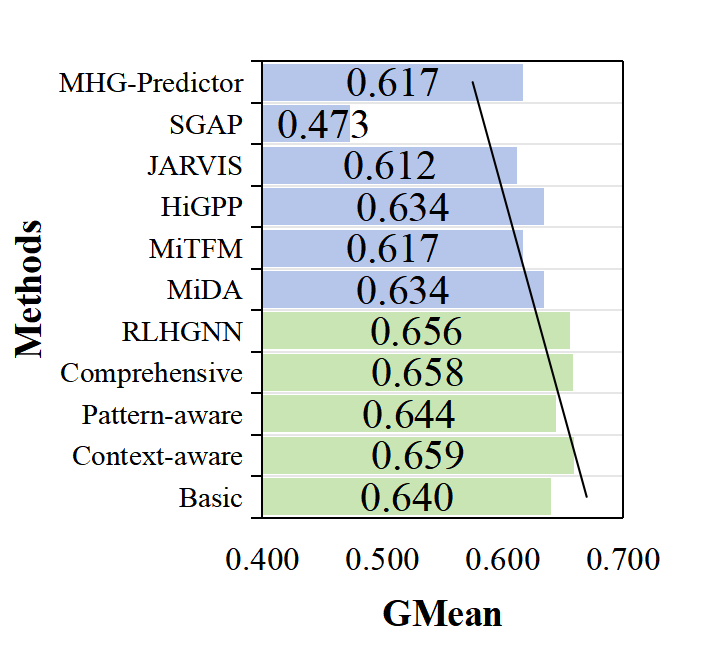}
        \caption{BPI13P}
        \label{BPI13P}
    \end{subfigure}
    \hfill
    \begin{subfigure}[b]{0.24\textwidth}
        \includegraphics[trim=0 0 20 0, clip,width=\textwidth]{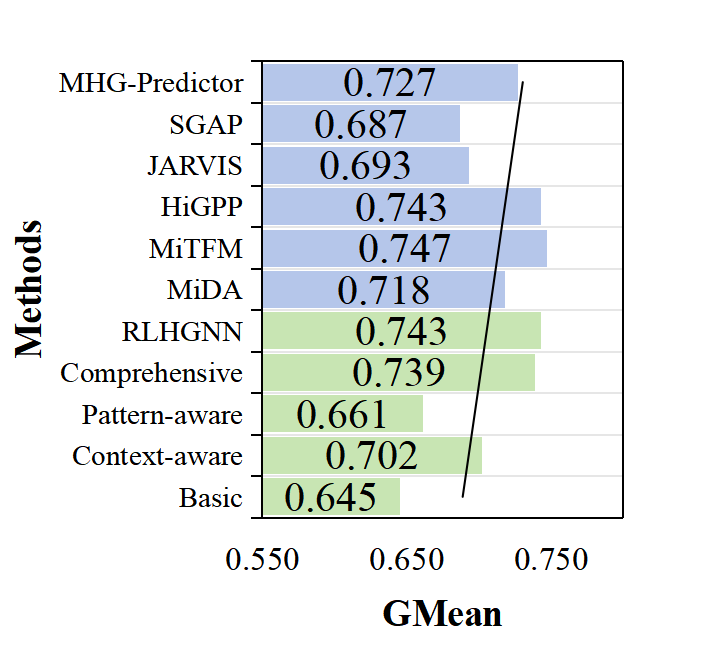}
        \caption{BPI2020P}
        \label{BPI2020P}
    \end{subfigure}
    \caption{GMean Comparison: RLHGNN, Fixed Graph Configurations, and Baseline Methods}
    \label{ablation-GMean}
\end{figure}

\begin{figure}
    \centering
    \begin{subfigure}[b]{0.24\textwidth}
        \includegraphics[trim=0 0 20 0, clip,width=\textwidth]{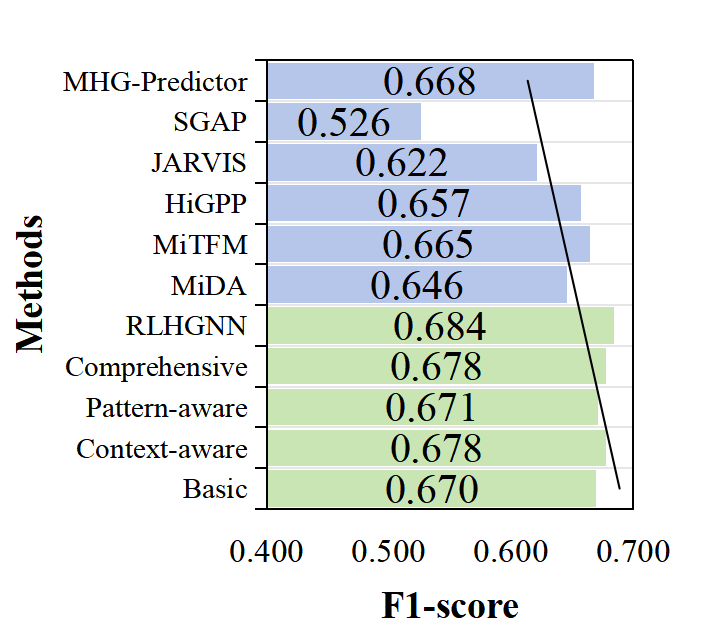}
        \caption{BPI12C}
        \label{BPI12C}
    \end{subfigure}
    \hfill
    \begin{subfigure}[b]{0.24\textwidth}
        \includegraphics[trim=0 0 20 0, clip,width=\textwidth]{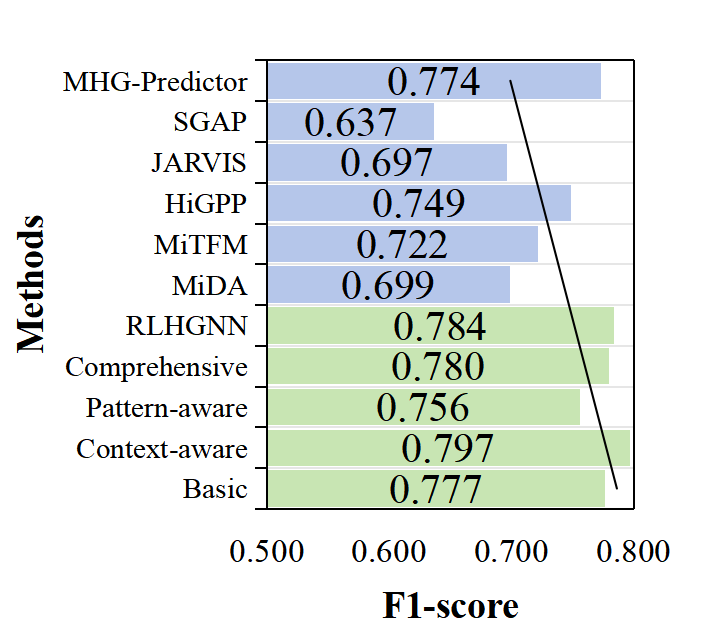}
        \caption{BPI12CW}
        \label{BPI12CW}
    \end{subfigure}
    
    \begin{subfigure}[b]{0.24\textwidth}
        \includegraphics[trim=0 0 20 0, clip,width=\textwidth]{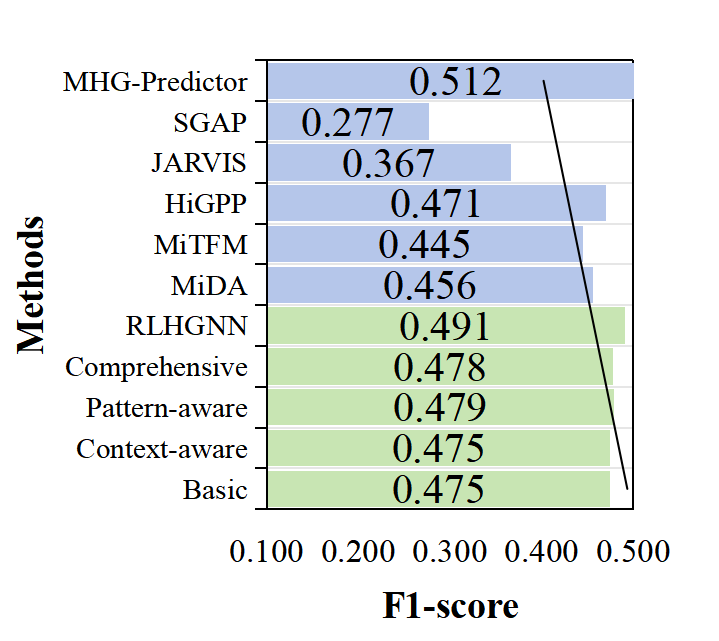}
        \caption{BPI13CP}
        \label{BPI13CP}
    \end{subfigure}
    \hfill
    \begin{subfigure}[b]{0.24\textwidth}
        \includegraphics[trim=0 0 20 0, clip,width=\textwidth]{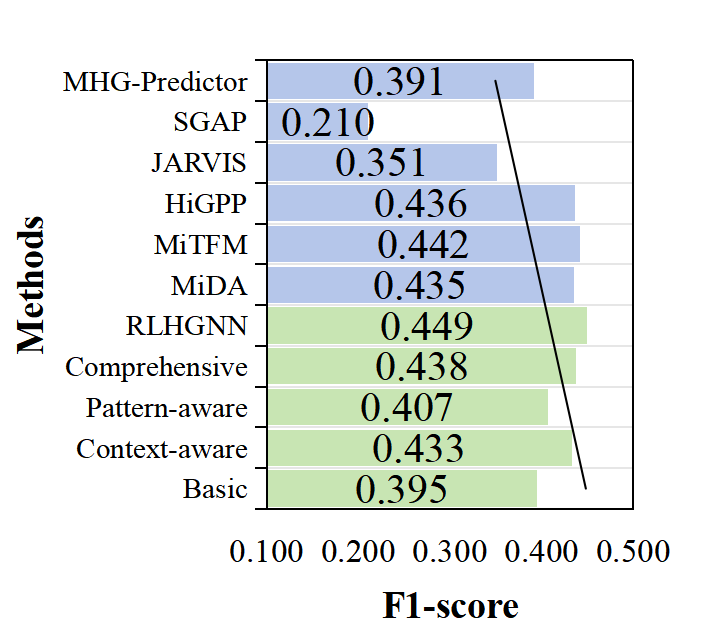}
        \caption{BPI13I}
        \label{BPI13I}
    \end{subfigure}

    \begin{subfigure}[b]{0.24\textwidth}
        \includegraphics[trim=0 0 20 0, clip,width=\textwidth]{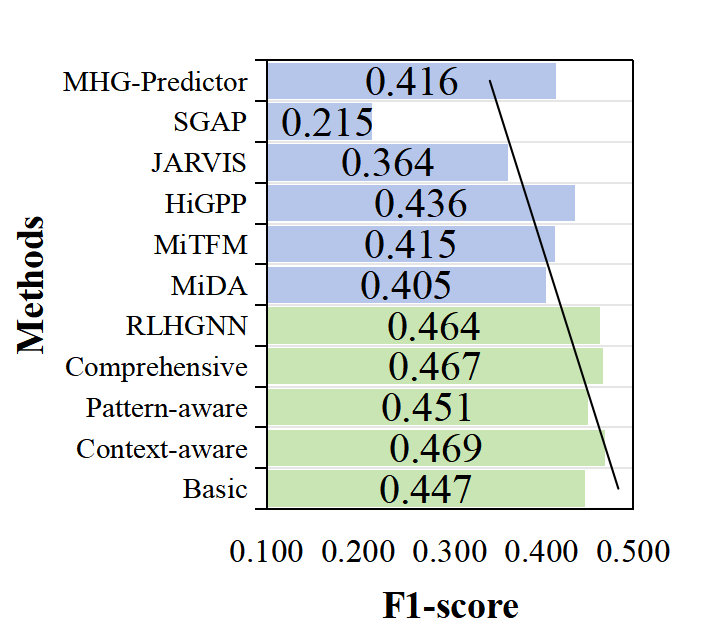}
        \caption{BPI13P}
        \label{BPI13P}
    \end{subfigure}
    \hfill
    \begin{subfigure}[b]{0.24\textwidth}
        \includegraphics[trim=0 0 20 0, clip,width=\textwidth]{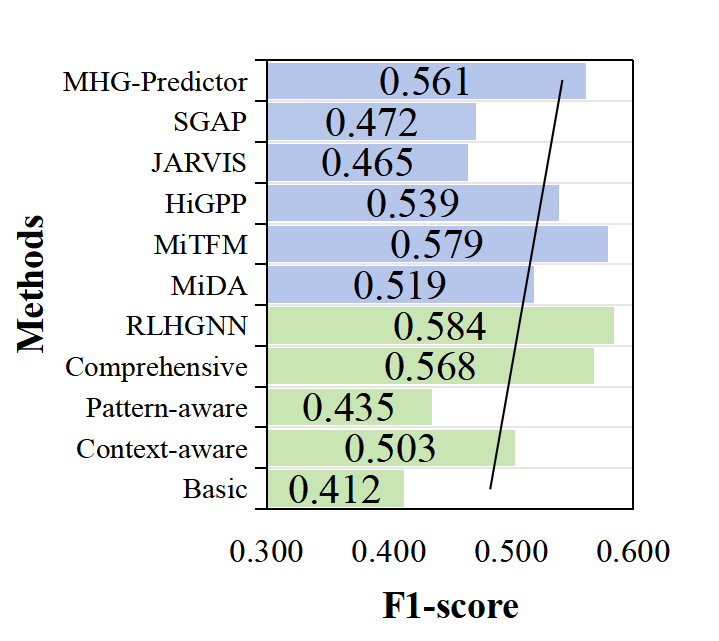}
        \caption{BPI2020P}
        \label{BPI2020P}
    \end{subfigure}
    \caption{F1-score comparison of RLHGNN, fixed graph structures, and baseline methods}
    \label{ablation-f1score}
\end{figure}

Fig. \ref{ablation-GMean} and Fig. \ref{ablation-f1score} provide visual comparisons across methods, demonstrating the effectiveness of our heterogeneous graph approach. Even our static graph structures achieve competitive performance—for instance, on BPI12CW, all static graph structures maintain GMean above 0.849, approaching the strongest baseline HiGPP (0.846). While RLHGNN does not achieve the highest score on every individual dataset, its value lies in consistent strong performance without manual configuration. On BPI12CW, the Context-aware structure achieves the best results (GMean: 0.876, F1-score: 0.797), slightly outperforming RLHGNN (0.872, 0.784). This occurs because BPI12CW exhibits highly uniform sequential patterns with consistent backward dependencies, making a fixed Context-aware structure optimal. However, RLHGNN demonstrates its adaptive advantage on structurally diverse datasets—achieving the highest F1-scores on BPI13CP (0.491 vs 0.479 for the best static graph structure) and BPI2020P (0.584 vs 0.568). The performance pattern validates our design principle: when process instances within a dataset vary significantly in complexity, the RL agent effectively selects appropriate structures for each instance, yielding superior overall performance. This adaptive capability proves particularly valuable in real-world service orchestrations where process complexity cannot be predetermined, making RLHGNN a robust solution for diverse business process contexts.

\subsection{Computational Efficiency Analysis} 
To answer RQ3, we evaluate RLHGNN's computational overhead by comparing training efficiency and inference latency against baseline methods across benchmark datasets. This evaluation examines both training time requirements and real-time inference performance to establish RLHGNN's computational positioning relative to existing approaches.

\begin{table*}[]
    \centering
    \makeatletter
    \patchcmd{\@makecaption}{\scshape}{}{}{}
    \caption{Total Training Time Comparison Across Next Activity Prediction Methods}
    \makeatother

\begin{tabular}{cccccccc}
\hline
\textbf{Unit: hour}  & \textbf{RLHGNN} & \textbf{MiDA}   & \textbf{MiTFM}                         & \textbf{HiGPP}                         & \textbf{JARVIS} & \textbf{SGAP}    & \textbf{MHG-Predictor}                 \\ \hline
BPI12C                & 6.512           & 207.105         & 0.860                                  & 3.836                                  & 21.535          & 285.346          & 4.197                                  \\
BPI12CW               & 3.208           & 41.027          & 0.354                                  & 1.445                                  & 6.481           & 141.482          & 1.617                                  \\
BPI13I                & 3.687           & 23.228          & 0.416                                  & 1.561                                  & 12.532          & 137.188          & 3.056                                  \\
BPI13P                & 2.146           & 2.206           & 0.047                                  & 0.253                                  & 1.722           & 10.991           & 0.390                                  \\
BPI13CP               & 1.791           & 1.835           & 0.035                                  & 0.133                                  & 0.929           & 8.523            & 0.256                                  \\
BPI2020P              & 2.404           & 11.595          & 0.144                                  & 0.423                                  & 2.879           & 38.845           & 1.216                                  \\ \hline
\textbf{Average Time} & \textbf{3.291}  & \textbf{47.833} & \cellcolor[HTML]{00B0F0}\textbf{0.309} & \cellcolor[HTML]{EE822F}\textbf{1.275} & \textbf{7.680}  & \textbf{103.729} & \cellcolor[HTML]{FFFF00}\textbf{1.789} \\ \hline
\end{tabular}

\captionof*{tablenote}{ 
\fontsize{9pt}{0pt}\selectfont
The best, second-best, and third-best performing approaches are highlighted in cyan, orange, and yellow, respectively.
}
\label{table:total_reasoning_time}

\end{table*}

\begin{table*}[]
    \centering
    \makeatletter
    \patchcmd{\@makecaption}{\scshape}{}{}{}
    \caption{Average Inference Latency Comparison per Prediction Instance}
    \makeatother

\begin{tabular}{cccccccc}
\hline
\textbf{Unit: hour}  & \textbf{RLHGNN}                        & \textbf{MiDA}                          & \textbf{MiTFM}                         & \textbf{HiGPP} & \textbf{JARVIS} & \textbf{SGAP}  & \textbf{MHG-Predictor} \\ \hline
BPI12C                & 0.810                                  & 0.382                                  & 0.086                                  & 4.319          & 0.611           & 0.540          & 1.890                  \\
BPI12CW               & 0.798                                  & 0.327                                  & 0.095                                  & 3.198          & 0.164           & 2.139          & 1.796                  \\
BPI13I                & 1.119                                  & 0.555                                  & 0.142                                  & 4.146          & 0.815           & 1.605          & 3.000                  \\
BPI13P                & 0.907                                  & 1.096                                  & 0.222                                  & 3.149          & 4.481           & 1.679          & 2.405                  \\
BPI13CP               & 1.171                                  & 1.149                                  & 0.153                                  & 3.186          & 1.056           & 1.960          & 3.048                  \\
BPI2020P              & 1.304                                  & 0.670                                  & 0.176                                  & 4.732          & 0.994           & 1.844          & 4.226                  \\ \hline
\textbf{Average Time} & \cellcolor[HTML]{FFFF00}\textbf{1.018} & \cellcolor[HTML]{EE822F}\textbf{0.696} & \cellcolor[HTML]{00B0F0}\textbf{0.146} & \textbf{3.788} & \textbf{1.354}  & \textbf{1.628} & \textbf{2.727}         \\ \hline
\end{tabular}

\captionof*{tablenote}{ 
\fontsize{9pt}{0pt}\selectfont
The best, second-best, and third-best performing approaches are highlighted in cyan, orange, and yellow, respectively.
}

\label{table:average_reasoning_time}
\end{table*}


 Tables \ref{table:total_reasoning_time} and \ref{table:average_reasoning_time} present comprehensive training time and inference latency comparisons across evaluated methods. RLHGNN requires 3.291 hours average training time, which positions it as the fourth most efficient approach among the seven methods evaluated. This moderate training overhead arises from the framework's sophisticated architectural design, which maintains four distinct graph structures and trains an RL agent to optimize structure selection. However, this training investment delivers substantial returns through superior predictive performance, with RLHGNN achieving the highest average accuracy (0.782) and F1-score (0.578) across all benchmark datasets. The performance advantage justifies the moderate training overhead through improved business process predictions that enable better resource allocation and service orchestration decisions. Furthermore, the training frequency context minimizes practical impact, as business process prediction models typically undergo retraining on monthly or quarterly cycles, so the additional training time represents minimal operational burden when amortized across extended deployment periods.
 
The inference efficiency analysis reveals an important trade-off between model complexity and runtime performance. RLHGNN's average inference latency of 1.018 milliseconds reflects the computational cost of dynamic graph structure selection and heterogeneous graph processing. While this represents a 7-fold increase compared to MiTFM (0.146ms), it remains well within practical bounds for real-time process monitoring applications. The efficiency patterns across datasets provide insights into RLHGNN's scalability characteristics. On simpler datasets like BPI13P and BPI13CP, RLHGNN maintains competitive inference times despite its adaptive architecture. However, on complex datasets with numerous activities like BPI2020P, the inference overhead becomes more pronounced (1.304ms), though still faster than several baselines. This suggests that RLHGNN's adaptive mechanism scales reasonably with process complexity, making it suitable for diverse business environments where process characteristics vary significantly.

\subsection{Aggregation Strategy Effectiveness Analysis}
To answer RQ4, we compare RLHGNN's heterogeneous aggregation approach against uniform aggregation strategies applied consistently across all edge types. This evaluation examines the effectiveness of relation-specific aggregation design versus uniform aggregation methods through performance analysis across multiple benchmark datasets.

\begin{figure}[h]
    \centering
    \begin{subfigure}[b]{1\linewidth}
        \centering
        \adjustbox{trim=0 0 0 0,clip}{\includegraphics[width=1\linewidth]{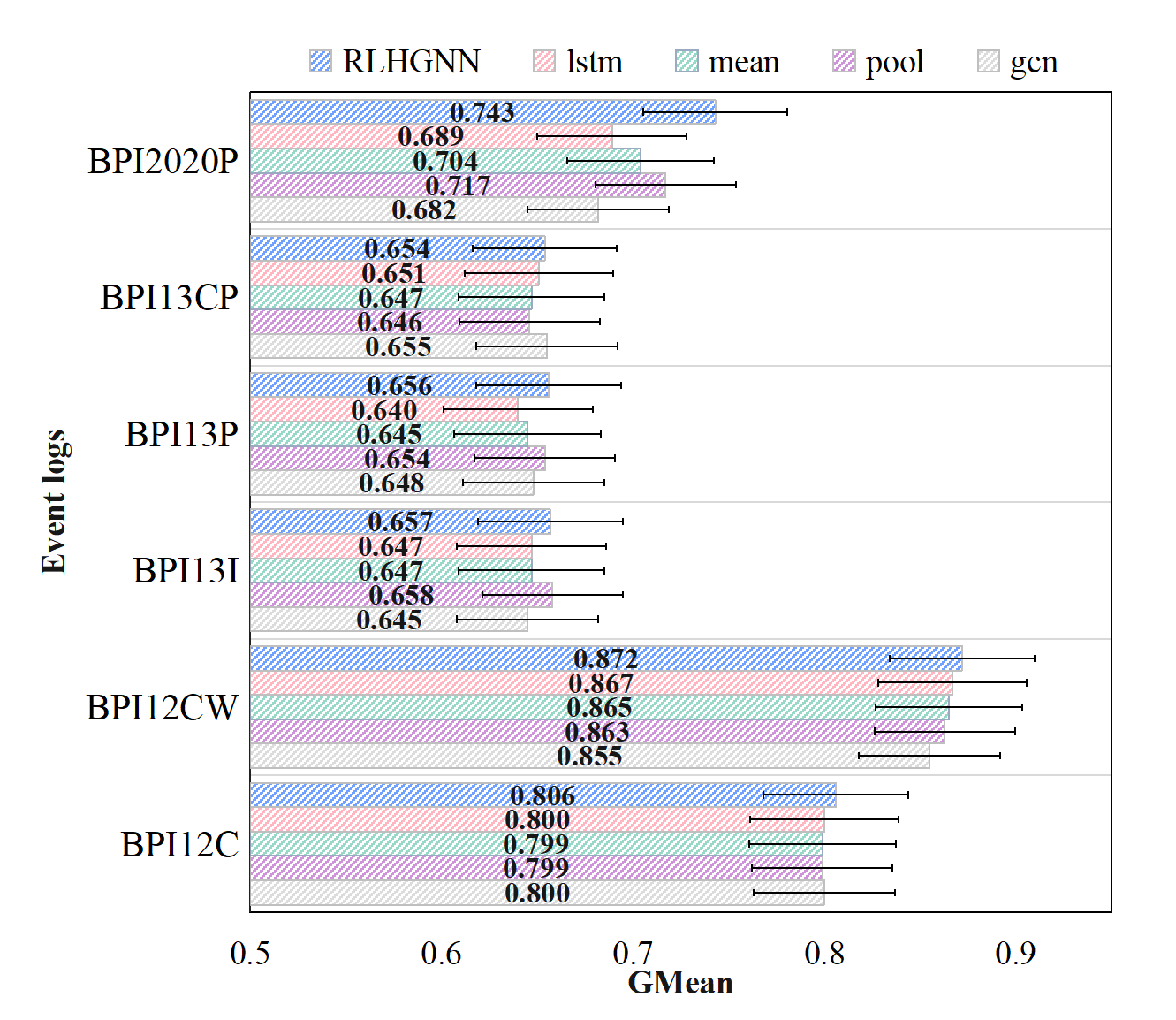}}
        \caption{GMean comparison across different aggregation strategies}
        \label{fig:aggregation-2}
    \end{subfigure}
    
    \begin{subfigure}[b]{1\linewidth}
        \centering
        \adjustbox{trim=0 0 0 0,clip}{\includegraphics[width=1\linewidth]{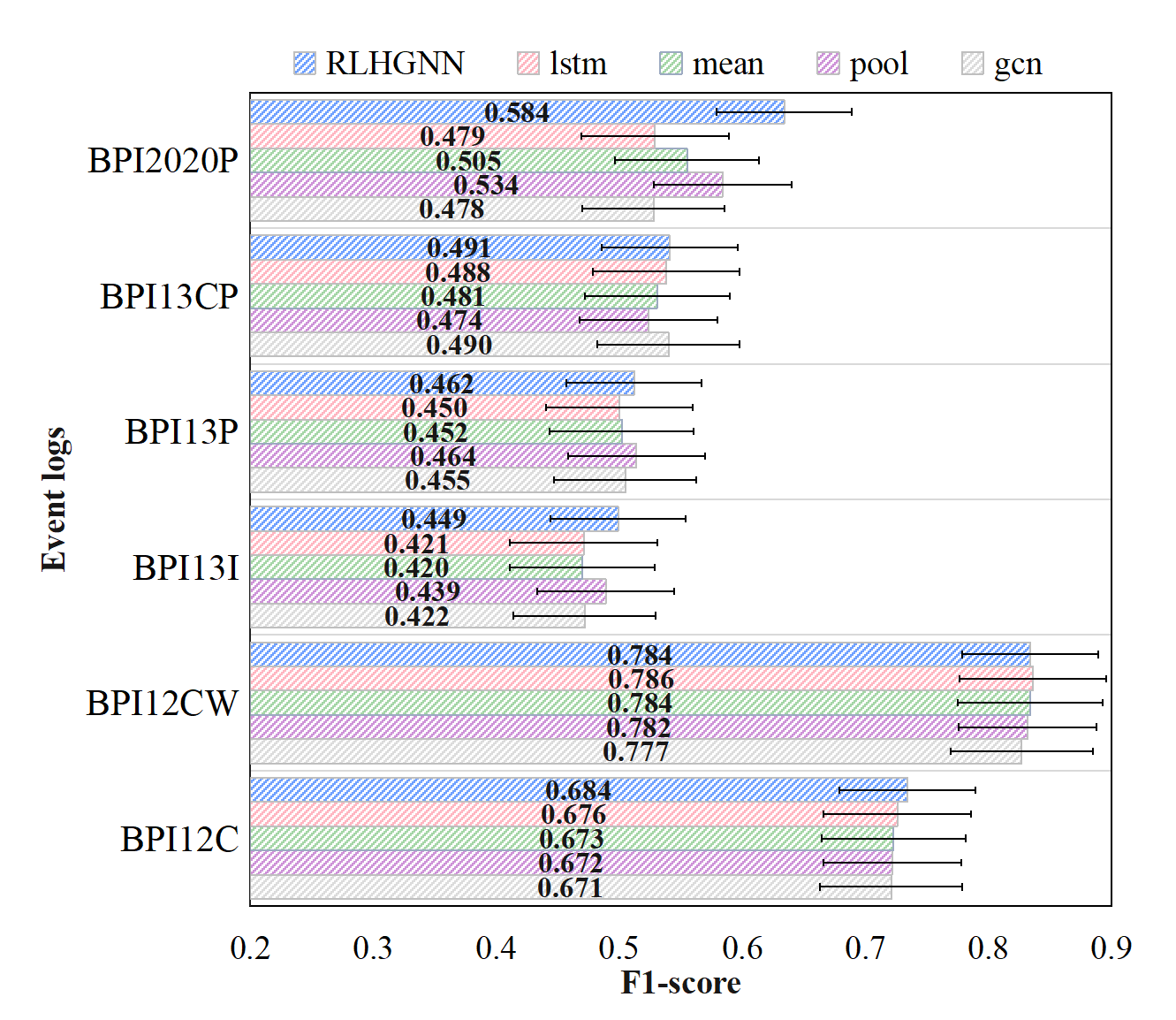}}
        \caption{F1-score comparison across different aggregation strategies}
        \label{fig:aggregation}
    \end{subfigure}
    
    \caption{Aggregation strategy analysis for GraphSAGE}
    \label{fig:aggregation_two}
\end{figure}

Fig. \ref{fig:aggregation_two} compares different aggregation strategies applied across all edge types versus RLHGNN's heterogeneous approach, evaluating both GMean and F1-score metrics. RLHGNN employs LSTM aggregation for forward and backward edges to preserve temporal dependencies, while using mean aggregation for repeat edges to capture pattern similarities. The experimental comparison includes RLHGNN alongside four uniform aggregation strategies: LSTM, Mean, Pool, and GCN, each applied uniformly across all edge types in the graph structure.

The experimental results demonstrate varying levels of sensitivity to aggregation strategy selection across different datasets. On BPI2020P, RLHGNN achieves consistent superior performance compared to all uniform aggregation strategies across both evaluation metrics, while the four uniform strategies show notable performance variations, indicating that aggregation method selection significantly impacts prediction effectiveness for this process type. Conversely, datasets such as BPI12C and BPI13CP demonstrate reduced sensitivity to aggregation strategy choice, with performance differences among uniform strategies remaining relatively modest while RLHGNN maintains competitive results. This pattern indicates that the impact of aggregation strategy selection varies considerably across different business process types.

The comparative analysis reveals that no uniform aggregation strategy achieves consistent superiority across all evaluated datasets. Strategies that perform well on one dataset may show diminished effectiveness on others, demonstrating the inherent challenge of selecting optimal aggregation approaches for diverse business process prediction scenarios. RLHGNN's relation-specific aggregation design addresses this variability by applying distinct methods tailored to each edge type based on semantic characteristics defined in the methodology.

This approach provides significant practical advantages for deployment scenarios where process characteristics cannot be predetermined or may vary over time. RLHGNN eliminates the need for dataset-specific aggregation strategy selection while maintaining robust performance across different process types. The consistent results across varying datasets position the method as suitable for production environments requiring reliable prediction performance without extensive domain-specific tuning or expertise in aggregation strategy optimization.

\section{Conclusion}
This paper presents RLHGNN, a novel framework combining RL with HGNN for next activity prediction in business processes. By introducing heterogeneous process graphs with three semantically distinct edge types grounded in process mining theory, combined with four adaptive graph structures and RL formulated as a Markov Decision Process for automatic structure selection, RLHGNN enables precise modeling of both sequential and non-sequential relationships in service interactions. This adaptive methodology addresses fundamental limitations of existing approaches that apply uniform modeling strategies regardless of process complexity diversity and lack automatic adaptation mechanisms to determine appropriate structural representations for different service interaction patterns. Comprehensive evaluation demonstrates the framework's effectiveness across diverse real-world datasets, achieving superior performance while maintaining computational efficiency for practical deployment. The adaptive selection mechanism provides robust prediction capabilities across varying process complexity characteristics, making it particularly valuable for service-oriented environments where process structures cannot be predetermined. 

Future work will enhance representation learning for sparse service interaction scenarios through service-oriented data augmentation techniques that leverage domain knowledge to generate synthetic process instances, which could improve policy learning effectiveness in data-scarce environments. Given that the current framework requires sufficient structural regularity to function effectively, we will also investigate hybrid approaches that combine adaptive selection with fallback mechanisms to static methods when insufficient training data prevents effective RL.

\bibliographystyle{IEEEtran}
\bibliography{sample}

\begin{thebibliography}{10}
\providecommand{\url}[1]{#1}
\csname url@samestyle\endcsname
\providecommand{\newblock}{\relax}
\providecommand{\bibinfo}[2]{#2}
\providecommand{\BIBentrySTDinterwordspacing}{\spaceskip=0pt\relax}
\providecommand{\BIBentryALTinterwordstretchfactor}{4}
\providecommand{\BIBentryALTinterwordspacing}{\spaceskip=\fontdimen2\font plus
\BIBentryALTinterwordstretchfactor\fontdimen3\font minus
  \fontdimen4\font\relax}
\providecommand{\BIBforeignlanguage}[2]{{%
\expandafter\ifx\csname l@#1\endcsname\relax
\typeout{** WARNING: IEEEtran.bst: No hyphenation pattern has been}%
\typeout{** loaded for the language `#1'. Using the pattern for}%
\typeout{** the default language instead.}%
\else
\language=\csname l@#1\endcsname
\fi
#2}}
\providecommand{\BIBdecl}{\relax}
\BIBdecl

\bibitem{liu2023discovering}
C.~Liu, Y.~Wang, L.~Wen, J.~Cheng, L.~Cheng, and Q.~Zeng, ``Discovering
  hierarchical multi-instance business processes from event logs,'' \emph{IEEE
  Transactions on Services Computing}, vol.~17, no.~1, pp. 142--155, 2023.

\bibitem{beheshti2023processgpt}
A.~Beheshti, J.~Yang, Q.~Z. Sheng, B.~Benatallah, F.~Casati, S.~Dustdar,
  H.~R.~M. Nezhad, X.~Zhang, and S.~Xue, ``Processgpt: transforming business
  process management with generative artificial intelligence,'' in \emph{2023
  IEEE International Conference on Web Services (ICWS)}.\hskip 1em plus 0.5em
  minus 0.4em\relax IEEE, 2023, pp. 731--739.

\bibitem{marques2024proactive}
G.~Marques, C.~Senna, S.~Sargento, L.~Carvalho, L.~Pereira, and R.~Matos,
  ``Proactive resource management for cloud of services environments,''
  \emph{Future Generation Computer Systems}, vol. 150, pp. 90--102, 2024.

\bibitem{polato2018time}
M.~Polato, A.~Sperduti, A.~Burattin, and M.~d. Leoni, ``Time and activity
  sequence prediction of business process instances,'' \emph{Computing}, vol.
  100, pp. 1005--1031, 2018.

\bibitem{sola2021rule}
D.~Sola, C.~Meilicke, H.~van~der Aa, and H.~Stuckenschmidt, ``A rule-based
  recommendation approach for business process modeling,'' in \emph{Advanced
  Information Systems Engineering: 33rd International Conference, CAiSE 2021,
  Melbourne, VIC, Australia, June 28--July 2, 2021, Proceedings}.\hskip 1em
  plus 0.5em minus 0.4em\relax Springer, 2021, pp. 328--343.

\bibitem{bohmer2018probability}
K.~B{\"o}hmer and S.~Rinderle-Ma, ``Probability based heuristic for predictive
  business process monitoring,'' in \emph{On the Move to Meaningful Internet
  Systems. OTM 2018 Conferences: Confederated International Conferences:
  CoopIS, C\&TC, and ODBASE 2018, Valletta, Malta, October 22-26, 2018,
  Proceedings, Part I}.\hskip 1em plus 0.5em minus 0.4em\relax Springer, 2018,
  pp. 78--96.

\bibitem{lakshmanan2015markov}
G.~T. Lakshmanan, D.~Shamsi, Y.~N. Doganata, M.~Unuvar, and R.~Khalaf, ``A
  markov prediction model for data-driven semi-structured business processes,''
  \emph{Knowledge and Information Systems}, vol.~42, pp. 97--126, 2015.

\bibitem{appice2019leveraging}
A.~Appice, N.~Di~Mauro, and D.~Malerba, ``Leveraging shallow machine learning
  to predict business process behavior,'' in \emph{2019 IEEE International
  Conference on Services Computing (SCC)}.\hskip 1em plus 0.5em minus
  0.4em\relax IEEE, 2019, pp. 184--188.

\bibitem{tama2019empirical}
B.~A. Tama and M.~Comuzzi, ``An empirical comparison of classification
  techniques for next event prediction using business process event logs,''
  \emph{Expert Systems with Applications}, vol. 129, pp. 233--245, 2019.

\bibitem{tax2017predictive}
N.~Tax, I.~Verenich, M.~La~Rosa, and M.~Dumas, ``Predictive business process
  monitoring with lstm neural networks,'' in \emph{Advanced Information Systems
  Engineering: 29th International Conference, CAiSE 2017, Essen, Germany, June
  12-16, 2017, Proceedings 29}.\hskip 1em plus 0.5em minus 0.4em\relax
  Springer, 2017, pp. 477--492.

\bibitem{camargo2019learning}
M.~Camargo, M.~Dumas, and O.~Gonz{\'a}lez-Rojas, ``Learning accurate lstm
  models of business processes,'' in \emph{Business Process Management: 17th
  International Conference, BPM 2019, Vienna, Austria, September 1--6, 2019,
  Proceedings 17}.\hskip 1em plus 0.5em minus 0.4em\relax Springer, 2019, pp.
  286--302.

\bibitem{sun2024next}
X.~Sun, S.~Yang, Y.~Ying, and D.~Yu, ``Next activity prediction of ongoing
  business processes based on deep learning,'' \emph{Expert Systems}, vol.~41,
  no.~5, p. e13421, 2024.

\bibitem{bukhsh2021processtransformer}
Z.~A. Bukhsh, A.~Saeed, and R.~M. Dijkman, ``Processtransformer: Predictive
  business process monitoring with transformer network,'' \emph{arXiv preprint
  arXiv:2104.00721}, 2021.

\bibitem{jalayer2022ham}
A.~Jalayer, M.~Kahani, A.~Pourmasoumi, and A.~Beheshti, ``Ham-net: Predictive
  business process monitoring with a hierarchical attention mechanism,''
  \emph{Knowledge-Based Systems}, vol. 236, p. 107722, 2022.

\bibitem{nguyen2024switch}
T.-H. Nguyen, K.-S. Kim, and K.~P. Kim, ``A switch-transformer predictive
  process monitoring model for next activity prediction in business process
  management,'' in \emph{Proceedings of the 2024 8th International Conference
  on Advances in Artificial Intelligence}, 2024, pp. 196--200.

\bibitem{zare2025innovative}
H.~Zare, M.~Abbasi, M.~Ahang, and H.~Najjaran, ``An innovative next activity
  prediction approach using process entropy and daw-transformer,'' \emph{arXiv
  preprint arXiv:2502.10573}, 2025.

\bibitem{chen2022multi}
H.~Chen, X.~Fang, and H.~Fang, ``Multi-task prediction method of business
  process based on bert and transfer learning,'' \emph{Knowledge-Based
  Systems}, vol. 254, p. 109603, 2022.

\bibitem{donadello2023knowledge}
I.~Donadello, J.~Ko, F.~M. Maggi, J.~Mendling, F.~Riva, and M.~Weidlich,
  ``Knowledge-driven modulation of neural networks with attention mechanism for
  next activity prediction,'' \emph{arXiv preprint arXiv:2312.08847}, 2023.

\bibitem{theis2022improving}
J.~Theis and H.~Darabi, ``Improving predictive process monitoring through
  reachability graph-based masking of neural networks,'' \emph{IEEE
  Transactions on Computational Social Systems}, vol.~10, no.~4, pp.
  1927--1938, 2022.

\bibitem{razo2023adjacency}
M.~Razo and H.~Darabi, ``Adjacency matrix deep learning prediction model for
  prognosis of the next event in a process,'' \emph{IEEE Access}, vol.~11, pp.
  11\,947--11\,955, 2023.

\bibitem{venugopal2021comparison}
I.~Venugopal, J.~T{\"o}llich, M.~Fairbank, and A.~Scherp, ``A comparison of
  deep-learning methods for analysing and predicting business processes,'' in
  \emph{2021 International Joint Conference on Neural Networks (IJCNN)}.\hskip
  1em plus 0.5em minus 0.4em\relax IEEE, 2021, pp. 1--8.

\bibitem{rama2023embedding}
E.~Rama-Maneiro, J.~C. Vidal, and M.~Lama, ``Embedding graph convolutional
  networks in recurrent neural networks for predictive monitoring,'' \emph{IEEE
  Transactions on Knowledge and Data Engineering}, vol.~36, no.~1, pp.
  137--151, 2023.

\bibitem{chiorrini2021exploiting}
A.~Chiorrini, C.~Diamantini, A.~Mircoli, and D.~Potena, ``Exploiting instance
  graphs and graph neural networks for next activity prediction,'' in
  \emph{International conference on process mining}.\hskip 1em plus 0.5em minus
  0.4em\relax Springer, 2021, pp. 115--126.

\bibitem{chiorrini2023multi}
A.~Chiorrini, C.~Diamantini, L.~Genga, and D.~Potena, ``Multi-perspective
  enriched instance graphs for next activity prediction through graph neural
  network,'' \emph{Journal of Intelligent Information Systems}, vol.~61, no.~1,
  pp. 5--25, 2023.

\bibitem{van2016process}
W.~M.~P. van~der Aalst, \emph{Process Mining: Data Science in Action},
  2nd~ed.\hskip 1em plus 0.5em minus 0.4em\relax Springer, 2016.

\bibitem{weijters2006process}
A.~J. M.~M. Weijters, W.~M.~P. van~der Aalst, and A.~K. Alves~de Medeiros,
  ``Process mining with the heuristics miner-algorithm,'' \emph{Technische
  Universiteit Eindhoven, Tech. Rep. WP}, vol. 166, 2006.

\bibitem{leemans2013discovering}
S.~J.~J. Leemans, D.~Fahland, and W.~M.~P. van~der Aalst, ``Discovering
  block-structured process models from event logs - a constructive approach,''
  in \emph{International Conference on Applications and Theory of Petri Nets
  and Concurrency}.\hskip 1em plus 0.5em minus 0.4em\relax Springer, 2013, pp.
  311--329.

\bibitem{pravilovic2014process}
S.~Pravilovic, A.~Appice, and D.~Malerba, ``Process mining to forecast the
  future of running cases,'' in \emph{New Frontiers in Mining Complex Patterns:
  Second International Workshop, NFMCP 2013, Held in Conjunction with ECML-PKDD
  2013, Prague, Czech Republic, September 27, 2013, Revised Selected Papers
  2}.\hskip 1em plus 0.5em minus 0.4em\relax Springer, 2014, pp. 67--81.

\bibitem{unuvar2016leveraging}
M.~Unuvar, G.~T. Lakshmanan, and Y.~N. Doganata, ``Leveraging path information
  to generate predictions for parallel business processes,'' \emph{Knowledge
  and Information Systems}, vol.~47, pp. 433--461, 2016.

\bibitem{le2017hybrid}
M.~Le, B.~Gabrys, and D.~Nauck, ``A hybrid model for business process event and
  outcome prediction,'' \emph{Expert Systems}, vol.~34, no.~5, p. e12079, 2017.

\bibitem{ferilli2019activity}
S.~Ferilli and S.~Angelastro, ``Activity prediction in process mining using the
  woman framework,'' \emph{Journal of Intelligent Information Systems},
  vol.~53, pp. 93--112, 2019.

\bibitem{gunnarsson2023direct}
B.~R. Gunnarsson, S.~vanden Broucke, and J.~De~Weerdt, ``A direct data aware
  lstm neural network architecture for complete remaining trace and runtime
  prediction,'' \emph{IEEE Transactions on Services Computing}, vol.~16, no.~4,
  pp. 2330--2342, 2023.

\bibitem{ni2023predictive}
W.~Ni, G.~Zhao, T.~Liu, Q.~Zeng, and X.~Xu, ``Predictive business process
  monitoring approach based on hierarchical transformer,'' \emph{Electronics},
  vol.~12, no.~6, p. 1273, 2023.

\bibitem{wang2023mitfm}
J.~Wang, C.~Lu, B.~Cao, and J.~Fan, ``Mitfm: A multi-view information fusion
  method based on transformer for next activity prediction of business
  processes,'' in \emph{Proceedings of the 14th Asia-Pacific Symposium on
  Internetware}, 2023, pp. 281--291.

\bibitem{pasquadibisceglie2022multi}
V.~Pasquadibisceglie, A.~Appice, G.~Castellano, and D.~Malerba, ``A multi-view
  deep learning approach for predictive business process monitoring,''
  \emph{IEEE Transactions on Services Computing}, vol.~15, no.~4, pp.
  2382--2395, 2022.

\bibitem{oved2025snap}
A.~Oved, S.~Shlomov, S.~Zeltyn, N.~Mashkif, and A.~Yaeli, ``Snap: semantic
  stories for next activity prediction,'' in \emph{Proceedings of the AAAI
  Conference on Artificial Intelligence}, vol.~39, no.~28, 2025, pp.
  28\,871--28\,877.

\bibitem{pasquadibisceglie2024jarvis}
V.~Pasquadibisceglie, A.~Appice, G.~Castellano, and D.~Malerba, ``Jarvis:
  Joining adversarial training with vision transformers in next-activity
  prediction,'' \emph{IEEE Transactions on Services Computing}, vol.~17, no.~4,
  pp. 1593--1606, 2024.

\bibitem{deng2024enhancing}
Y.~Deng, J.~Wang, C.~Wang, C.~Zheng, M.~Li, and B.~Li, ``Enhancing predictive
  process monitoring with sequential graphs and trace attention,'' in
  \emph{2024 IEEE International Conference on Web Services (ICWS)}.\hskip 1em
  plus 0.5em minus 0.4em\relax IEEE, 2024, pp. 406--415.

\bibitem{wang2025higpp}
J.~Wang, C.~Lu, Y.~Yu, B.~Cao, K.~Fang, and J.~Fan, ``Higpp: A history-informed
  graph-based process predictor for next activity,'' in \emph{International
  Conference on Service-Oriented Computing}.\hskip 1em plus 0.5em minus
  0.4em\relax Springer, 2025, pp. 337--353.

\bibitem{wang2025mhg}
J.~Wang, Y.~Yu, N.~Fang, B.~Cao, J.~Fan, and J.~Zhang, ``Mhg-predictor: A
  multi-layer heterogeneous graph-based predictor for next activity in complex
  business processes,'' in \emph{Companion Proceedings of the ACM on Web
  Conference 2025}, 2025, pp. 500--509.

\bibitem{mnih2015human}
V.~Mnih, K.~Kavukcuoglu, D.~Silver, A.~A. Rusu, J.~Veness, M.~G. Bellemare,
  A.~Graves, M.~Riedmiller, A.~K. Fidjeland, G.~Ostrovski \emph{et~al.},
  ``Human-level control through deep reinforcement learning,'' \emph{nature},
  vol. 518, no. 7540, pp. 529--533, 2015.

\bibitem{shakya2023reinforcement}
A.~K. Shakya, G.~Pillai, and S.~Chakrabarty, ``Reinforcement learning
  algorithms: A brief survey,'' \emph{Expert Systems with Applications}, vol.
  231, p. 120495, 2023.

\bibitem{hamilton2017inductive}
W.~Hamilton, Z.~Ying, and J.~Leskovec, ``Inductive representation learning on
  large graphs,'' \emph{Advances in neural information processing systems},
  vol.~30, 2017.

\end{thebibliography}

\end{document}